\documentclass[10pt,final,journal]{IEEEtran}
\usepackage{cite}
\usepackage{amsmath,amssymb,amsfonts}
\usepackage{graphicx}
\usepackage{float}
\usepackage{subfigure}
\usepackage{mathrsfs}
\usepackage{threeparttable}
\usepackage{multirow}

\usepackage{color}

\ifCLASSINFOpdf
\else
\fi

\begin{document}
\graphicspath{{figures/}}
\title{Local Patch Network with Global Attention for Infrared Small Target Detection}

\author{    
 Fang Chen,
 Chenqiang Gao*,
 Fangcen Liu,
 Yue Zhao,
 Yuxi Zhou,
 Deyu Meng,~\IEEEmembership{Member,~IEEE},
 Wangmeng Zuo,~\IEEEmembership{Senior Member,~IEEE}% <-this % stops a space
\thanks{*Corresponding author: Chenqiang Gao.}% <-this % stops a space
 
% 致谢基金
% \thanks{This work is supported in part by the National Natural Science Foundation of China under Grant 61906025, in part by the Chongqing Research Program of Basic Research and Frontier Technology under and Grant cstc2020jcyj-msxmX0835, and in part by the Science and Technology Research Program of Chongqing Municipal Education Commission under Grant KJQN201900607 and Grant KJQN202000647.}% <-this % stops a space

\thanks{Fang Chen, Chenqiang Gao, Fangcen Liu, Yue Zhao, Yuxi Zhou are with the School of Communication and Information Engineering, Chongqing University of Posts and Telecommunications, and also with Chongqing Key Laboratory of Signal and Information Processing, Chongqing University of Posts and Telecommunications, Chongqing 400065, China (e-mail: cfun.cqupt@outlook.com, gaocq@cqupt.edu.cn, liufc67@gmail.com, zhaoyue@cqupt.edu.cn, 26680948@qq.com).}% <-this % stops a space
 
\thanks{Deyu Meng is with Macau Institute of Systems Engineering, Macau University of Science and Technology, Taipa, 999078, Macau and also with School of Mathematics and Statistics, Xi’an Jiaotong University, Xi’an, Shanxi, 710049, China (e-mail: dymeng@mail.xjtu.edu.cn).}% <-this % stops a space
 
\thanks{Wangmeng Zuo is with School of Computer Science and Technology, Harbin Institute of Techonlogy, 47822 Harbin, Heilongjiang, China (e-mail: wmzuo@hit.edu.cn).}% <-this % stops a space

% \thanks{Manuscript received April 19, 2005; revised September 17, 2014.}

}


\markboth{Local Patch Network with Global Attention for Infrared Small Target Detection}%
{Chen \MakeLowercase{\textit{et al.}}: Local Patch Network with Global Attention for Infrared Small Target Detection}



\maketitle

\begin{abstract}
 Infrared small target detection plays an important role in the infrared search and tracking applications.
 In recent years, deep learning techniques were introduced to this task and achieved noteworthy effects.
 Following general object segmentation methods, existing deep learning methods usually processed the image from the global view.
 However, the imaging locality of small targets and extreme class-imbalance between the target and background pixels were not well-considered by these deep learning methods, which causes the low-efficiency on training and high-dependence on numerous data.
 A local patch network (LPNet) with global attention is proposed in this paper to detect small targets by jointly considering the global and local properties of infrared small target images.
 From the global view, a supervised attention module trained by the small target spread map is proposed to suppress most background pixels irrelevant with small target features.
 From the local view, local patches are split from global features and share the same convolution weights with each other in a patch net.
 By leveraging both the global and local properties, the data-driven framework proposed in this paper has fused multi-scale features for small target detection.
 Extensive synthetic and real data experiments show that the proposed method achieves the state-of-the-art performance compared with existing both conventional and deep learning methods.
\end{abstract}


% \fi
\IEEEpeerreviewmaketitle

\section{Introduction}
\label{section:introduction}
\begin{figure}[htbp]
\centerline{\includegraphics[scale=0.262]{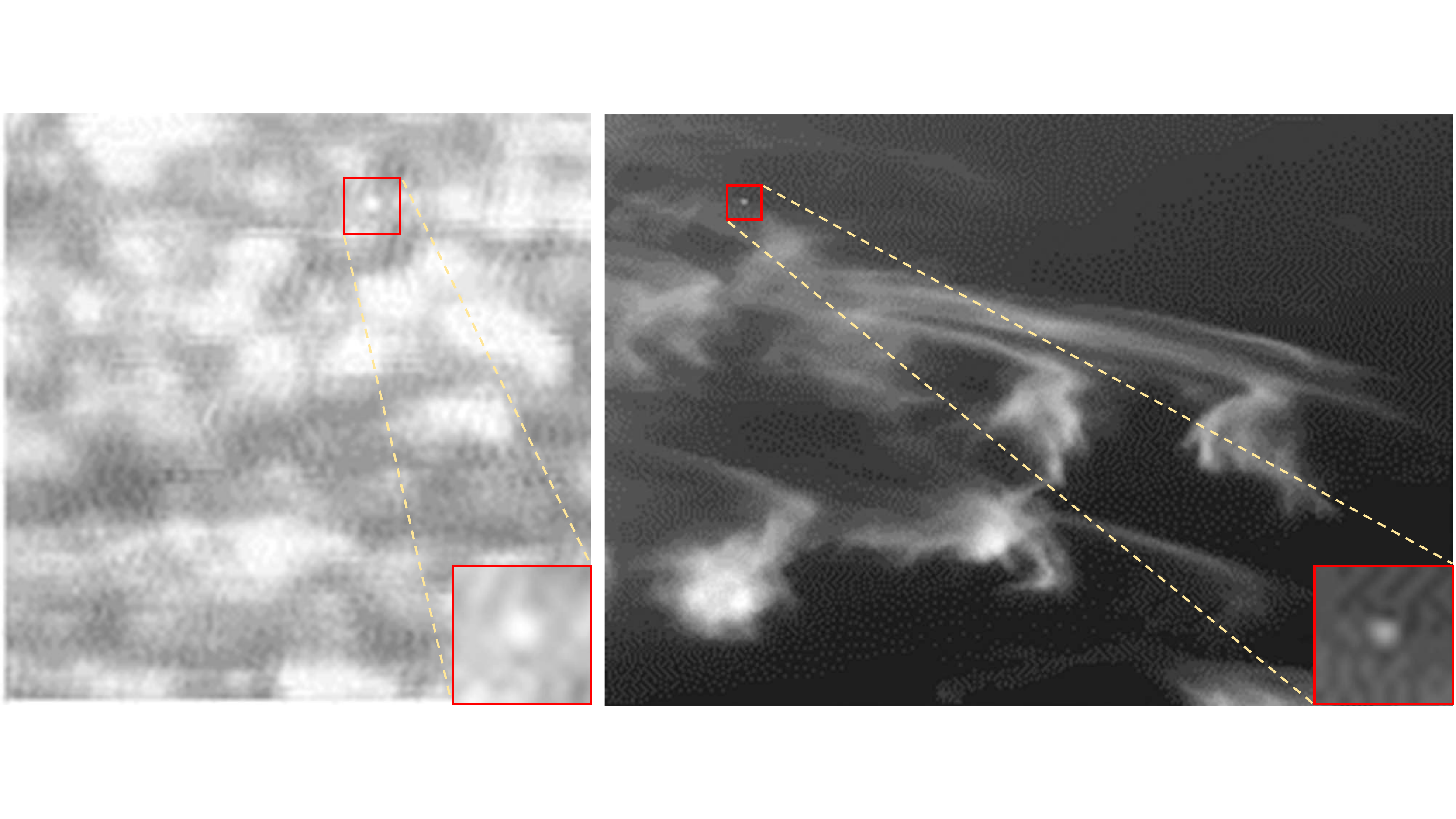}}
\caption{
 Representative examples of infrared small targets.
 The target is indicated by a red bounding box and a close-up is shown in the bottom right corner of each example.
 Left: an infrared small target image which has heavily cloudy clutters interfering detectors.
 Right: a target so small to lose its shape characteristics due to the long imaging distance.
 }
\label{fig:infrared small target detection}
\end{figure}

 % 介绍问题
 The infrared search and tracking (IRST) system plays an important role in kinds of applications, including early warning and maritime surveillance, etc.
 As one of critical techniques of IRST, infrared small target detection determines the performance of the IRST system and is still a challenging task since the target lacks obvious shape and texture characteristics.
 Besides, complex background clutters usually seriously interfere the detector, as shown in Fig.~\ref{fig:infrared small target detection}.
 
 In early stages, some methods based on background estimation~\cite{bai2010analysis, zeng2006design} were proposed through subtracting the estimated background from the original infrared image to detect small targets.
%  Some other methods based on denoising filter~\cite{gao2008generalised, Gao2018Infrared} regards the small target or background as noise and transforms this issue into a denoising task.
%  These methods were fast and had low computational complexity but always had a high false-alarm (FA) rate due to imperfect background estimation~\cite{moradi2020fast}.
 These methods were fast and had low computational complexity.
 Some other methods based on local contrast measure~\cite{aghaziyarati2019small, moradi2020fast, 8922738, xia2019infrared, zhang2018infrared} focused on target enhancement and background suppression through capturing local contrast and saliency.
%  These methods could effectively suppress background clutters, but the targets were also possibly filtered out from original images due to the low signal-to-clutter ratio (SCR) in local regions.
 These methods could effectively suppress clutters, especially when facing to complicated backgrounds.
 Some methods~\cite{gao2013infrared, Gao2018Infrared, wang2018infrared, zhang2019infrared1, zhang2019infrared2} based on nonlocal self-correlation property achieved favorable performance by transforming the task into a low-rank and sparse matrix separation problem.
%  However, the assumption of non-local self-correlation property does not always hold for kinds of scenes.
%  For some complex scenes, the performance of these methods could degrade.
 
 Totally, above methods fully rely on prior knowledge to design filters or models for small target detection.
 These methods are not feasible when applying to kinds of specific applications.
 When the prior knowledge of designing methods does not match that of a specific application scene, the new prior knowledge of the scene is not easily or fast embedded into these methods.
 In contrast, learning based methods with training samples are more feasible because the new prior knowledge embedding can be easily implemented through training samples from the scene.
 Recently, some deep learning frameworks~\cite{zhao2020novel, wang2019miss, zhao2019tbc} are proposed to detect infrared small targets by training models on large amounts of data.
 These methods are completely driven by data instead of expert knowledge so that better robustness and generalization have been verified.
 However, existing deep learning methods are based on the global view to extract features from infrared images and ignore the imaging locality of small targets.
 Only focusing on the global view causes numerous parameters of models and low-efficiency on training because the designed models have to exclude the disturbance of background which took up almost all pixels of an infrared image.
 Besides, the complexity and disparity of the background lead to these global view based methods depending on numerous training data with heterogeneous scenes in order to learn sufficient features for suppressing the background.
%  In contrast, in this paper the proposed method considers the local property of small targets and synthesizes the global and local features by a designed framework that gradually views the infrared image from the whole to the local.

 % 引入自己的方法并介绍贡献
 When just considering the global property of small target images, the class-imbalance of target and background pixels is the main obstacle for training a favorable and robust detector.
 Taking up only a few pixels, small targets are generally too small to be distinguished from the background.
 For example, in a $120\times120$ image containing a $2\times2$ small target, these four target pixels have to share the same convolution weights with the rest $14396$ background pixels.
 This will cause the extreme class-imbalance: the amount of small target pixels is much less than background pixels, which leads to performance degradation of deep learning frameworks.
 
 In this paper, we propose a local patch network (LPNet) with global attention for infrared small target detection from the global view and the local view.
 From the global view, we utilizes a small target spread map to train an attention module.
 This spread map and attention mechanism are designed to suppress distal background pixels irrelevant with the small target.
 By this way, most background pixels could be suppressed while the local saliency between small target pixels and adjacent background pixels is retained, so that the class-balance between small target pixels and background pixels is achieved.
 Besides, with most background pixels being suppressed, the complexity of background is greatly reduced and thus the proposed method could still effectively extract small target features even being trained on limited data with heterogeneous scenes.

 From the local view, the global feature maps are divided to a series of patches by using a sliding window to extract sub-images and a neural network is utilized to extract small target features and allocate a patch-likelihood map for each local patch.
 The ratio of small target pixels to background pixels in each patch is much higher than in the original image, so that the class-balance of small target and background is achieved.
 Besides, with sharing the same convolution weights for all local patches, the amount of parameters of the neural network is limited.
 
 To leverage both the global view to the local view, a well-designed framework is proposed to combine above modules for the global and local feature fusion.
 The proposed method is performed on two widely used public infrared small target datasets and a private dataset collected by ourselves to compare with state-of-the-art methods on comprehensive scenes.
 Besides, a series of ablation studies are performed for illustrating the remarkable effect of focusing on locality and class balance for small target detection.

 In summary, contributions of this paper can be summarized as follows:
 \begin{itemize}
     \item We provide a paradigm of deep learning methods to leverage both global and local properties of the infrared image for detecting small targets.
     \item We propose a supervised attention module trained on a small target spread map.
     This module suppresses background likely pixels and enhances small target likely pixels to force the whole network to focus on some local regions, which is effective to address the class-imbalance and reduce the complexity of the background.
     Besides, this attention module enables the proposed method to still maintain its effect even being trained on limited data with heterogeneous scenes.
     \item We propose a patch net to effectively extract local features correlated with small targets and fusing all these local features to provide the final predict.
     By patch splitting, the class-imbalance of target pixels and background pixels is further addressed.
     Through sharing the same convolution weights for all local patches, the task of small target detection is greatly simplified and the amount of parameters in patch net is limited.
     \item Extensive experiments on widely used and our private infrared small target datasets illustrate that the proposed method has achieved the best performance compared with state-of-the-art methods.
     Besides, our method has the ability of effectively extracting small target features even being trained on limited data with heterogeneous scenes.
 \end{itemize}

 The remainder of this paper is organized as follows:
 In Section~\ref{sec:related work}, related works are briefly reviewed.
 In Section~\ref{sec:the proposed method}, we present the proposed method in detail.
 In Section~\ref{sec:experiments and analysis}, the experimental results are given and discussed.
 Conclusions are drawn in Section~\ref{sec:conclusion}.
 
\section{Related work}
\label{sec:related work}
 Up to now, many methods for infrared small target detection have been proposed, which can be roughly categorized into two groups: conventional methods and deep learning methods.
 
 \textbf{Conventional methods} were instructed by prior knowledge to design filters or modules~\cite{shang2016infrared, deng2016infrared, liu2018tiny, wu2020double, zhou2020robust, li2018robust, bai2010analysis, deng2018adaptive, han2020infrared}.
 For example, some methods focused on the background estimation in early stages, such as the top-hat algorithm~\cite{zeng2006design} and max-mean/max-median algorithm~\cite{deshpande1999max}, which directly subtracted the estimated background from the original image to acquire detected small targets.
 The average absolute gray difference (AAGD) algorithm~\cite{aghaziyarati2019small}, absolute directional mean difference (ADMD) algorithm~\cite{moradi2020fast} and generalised structure tensor (GST) based algorithm~\cite{gao2008generalised} detected small targets through suppressing the estimated background as much as possible.
 Unlike these background estimation methods, some noise estimation methods regarded the dim-small target as noise and addressed the small target detection issue by estimating the noise.
 Gao et al.~\cite{Gao2018Infrared} proposed a Mixture of Gaussians (MoG) to model the small target as a special sparse noise component of the background noise by MoG with Markov random field (MRF), so that the targets could be separated from background by variational Bayesian.
%  These estimation methods had favorable performance on simple instances but were sensitive to noise or imperfect background estimation~\cite{kim2012scale} and could not perform well on complicated scenes, such as low SCR.

 Some local contrast methods designed special local filters to construct pixel-contrast in local regions.
 For example, local contrast measure (LCM)~\cite{chen2013local} and the tri-layer local contrast method (TLLCM)~\cite{8922738} applied an internal sliding window over the entire image to capture the local contrast and used an adaptive threshold to segment small targets from the image.
 Inspired by LCM and human visual system, some local based methods have been proposed~\cite{deng2016small, shi2017high, qin2016effective, 8660588, 8754801, 8705367}.
 Han et al.~\cite{han2014robust} proposed an improved local contrast measure (ILCM) to enhance the local saliency of small targets.
 Wei et al.~\cite{wei2016multiscale} proposed a multi-scale patch-based contrast measure (MPCM), which divided each image patch into nine cells and calculated the dissimilarities between the surrounding cells and the central one.
 By multiplying directional dissimilarities and minimum selection among the results, the final output was obtained.
 Based on MPCM, Xia et al.~\cite{xia2019infrared} conceived a local dissimilarity descriptor encoded by the local energy factor (LEF) and applied the LEF in MPCM.
 Gao et al.~\cite{8663277} applied a temporal variance filter after MPCM to remove small broken cloud regions and suppress noise.
 Zhang et al.~\cite{zhang2018infrared} proposed a method based on local intensity and gradient (LIG) properties of small targets to suppress clutters and enhance small targets.
%  These methods had low computational complexity.
%  However, the special local filters were based on prior knowledge and could not be general for covering all local scenes.
 
 Nonlocal self-correlating methods utilized the nonlocal self-correlating property of infrared background images and achieved favorable performance by transforming the task of detecting the small target into a low-rank and sparse matrix separation problem~\cite{zhang2018infrared, deng2018low, guan2020infrared, 8832236, ZHANG201955}.
 Gao et al. ~\cite{gao2013infrared} proposed the patch-image in infrared small target detection and constructed the low-rank based infrared patch-image (IPI) model.
 Dai et al.~\cite{dai2016infrared} proposed a column weighted IPI model (WIPI), and then proposed a re-weighted infrared patch-tensor model (RIPT)~\cite{dai2017reweighted}.
 Wang et al.~\cite{wang2018infrared} proposed a patch image model with local and global analysis (PILGA) to constrain the sparsity of noise patch images.
 Based on the IPI model, Wang et al.~\cite{wang2017infrared} mapped the background to multiple sub-spaces and proposed a low-rank and sparse decomposition method based on greedy bilateral factorization.
 %Zhang et al.~\cite{} added a post filtering module to IPI model for adaptive segmentation of small targets.
 There are also some methods constrained sparse term to suppress background or enhance target.
 Zhang et al.~\cite{zhang2019infrared1} proposed a novel infrared small target detection method based on non-convex optimization with $l_p$-norm constraint.
 Besides, Zhang et al.~\cite{zhang2019infrared2} employed a novel non-convex low-rank constraint named partial sum of tensor nuclear norm (PSTNN) joint weighted $l_1$ norm to efficiently suppress the background and preserve the target.
%  These methods had achieved good results in complicated scenes, but they suffered a high computational complexity~\cite{pang2020infrared}.
 
 In conclusion, the performance of these conventional methods were based on and limited by the expert prior on the characteristics of infrared small targets.
 However, due to the prior knowledge that these methods highly relied on, the conventional methods lacked generalization for detecting infrared small targets beyond the prior.
 A conventional method usually works well for a specific application, with limited generalization to other applications.
 Data driven is a potential technique to address this issue, which enables to learn features of infrared small targets from numerous samples, so that the generalization and effectiveness of detecting unknown infrared small targets beyond the prior are ensured.
 
 \textbf{Deep learning methods} recently have attracted more and more attention on infrared small target detection~\cite{simonyan2014very, he2016deep, lee2019energy, xie2016multispectral}.
 Some works proposed deep learning frameworks decoupled with prior knowledge and experts analysis~\cite{fan2018dim, zhao2019tbc, wang2019miss, shi2019infrared, zhao2020novel, ju2021istdet, hou2021ristdnet, ryu2019heterogeneous, gao2019dim, du2021cnn}.
 Specifically, these methods used global features of the infrared image extracted by designed neural networks to allocate a small target probability for each pixel.
 Wang et al.~\cite{wang2019miss} focused on balancing the Miss Detection (MD) and False Alarm (FA), using two sub-tasks handled by two-stream models trained adversarially, with each stream focusing on reducing either MD or FA.
 Ju et al.~\cite{ju2021istdet} proposed an end-to-end network named ISTDet to detect the dim and small target. 
 Motivated by the fact that the infrared small target has its unique distribution characteristics, Shi et al.~\cite{shi2019infrared} treated infrared small target as “noise” and transformed small target detection task into denoising problem.
 Zhao et al.~\cite{zhao2020novel} constructed a generative adversarial networks (GAN) model to learn the features of targets and directly predicted the intensity of targets, automatically.
%  To mitigate the issue of minimal intrinsic features for pure data-driven methods, Dai et al.~\cite{dai21acm} proposed a novel data-driven deep network for exchanging high-level semantics and subtle low-level details of the infrared small target detection.
 However, due to focusing on global features, these methods had to be assembled with numerous parameters for effectively extracting small target features from global features and highly relied on large amounts of data with discrepant scenes for learning features as much as possible.
%  These deep learning methods achieved both low computational complexity and high detecting precision.

 Compared with these deep learning methods, the proposed method additionally explores local property of the infrared small target, which is conductive to simplify the task on extracting small target features and could effectively extract small target features from limited data with various scenes.
%  Besides, through the supervised attention module and patch net, the complexity of background is greatly reduced and the class-imbalance issue between target pixels and background pixels is addressed.
%  This enables the proposed method to extract effective small target features from limited data with various scenes.

\section{The Proposed Method}
\label{sec:the proposed method}

\begin{figure*}[htbp]
\centerline{\includegraphics[scale=0.544]{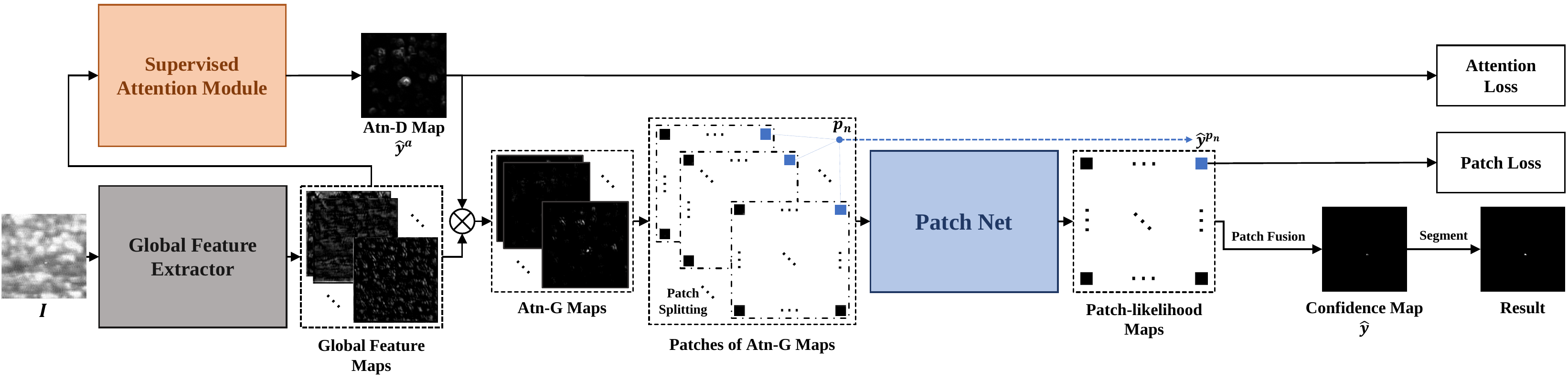}}
\caption{
 The architecture of the proposed method.
 The global feature extractor aims to extract basic features of the input infrared image.
 The supervised attention module suppresses inconsequential distant background pixels to force the network to focus on the local saliency of small targets.
 The patch net utilizes a designed convolutional neural network (CNN) to extract multi-scale features from each patch.
%  patch fusion is used to fuse all patch-likelihood maps to get the final likelihood map.
 }
\label{fig:architecture of the whole network}
\end{figure*}

 The overall architecture of the proposed method is shown as Fig.~\ref{fig:architecture of the whole network}.
 It mainly consists of three modules: global feature extractor, supervised attention module and patch net.
 The global feature extractor is used to extract basic features of the input image $I$ by viewing the whole image.
 Capturing these basic features is useful for reducing the complexity of the background.
 An attention-density (Atn-D) map $\hat{y}^{a}$ is calculated by the supervised attention module for the background suppression and the small target enhancement.
 The Atn-D map is applied on the global feature maps to get attention-global (Atn-G) maps.
 Based on a series of patches $p_n$ split from the Atn-G maps, local features correlated with small targets are extracted by the patch net to calculate patch-likelihood maps $\hat{y}^{p_n}$.
 By sharing the same convolution weights for all local patches, the task of small target detection is greatly simplified and the amount of parameters in the patch net can be limited.
 In each patch-likelihood map $\hat{y}^{p_n}$, the value of a pixel represents the probability that the pixel belongs to the background or the small target.
 Finally, a patch fusion strategy is applied to fuse these patch-likelihood maps and to get the final detection result by applying threshold segmentation on the fused likelihood map, i.e., confidence map $\hat{y}$.

\subsection{Global Feature Extractor}
\label{subsec:global feature extractor}

\begin{figure}[htbp]
\centerline{\includegraphics[scale=0.43]{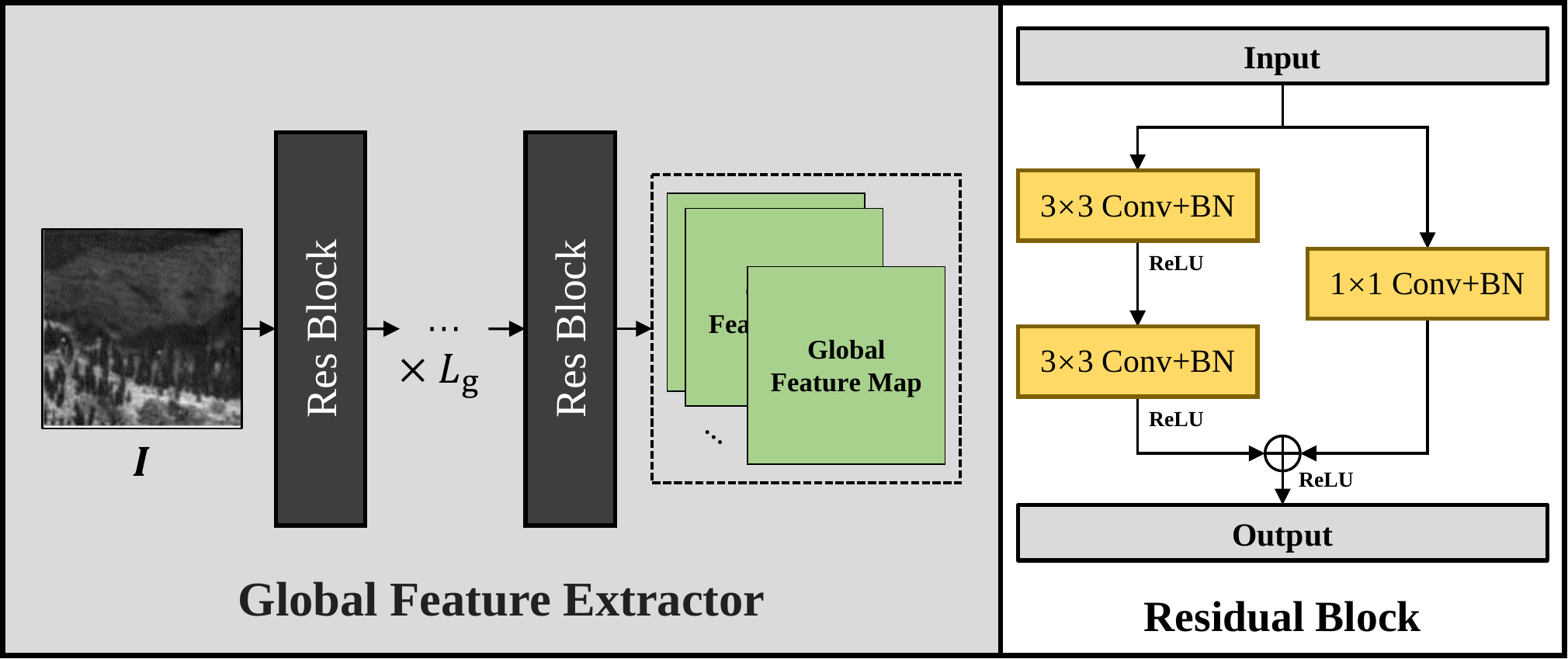}}
\caption{
 The architecture of the global feature extractor and the residual block.
}
\label{fig:architecture of GFE and residual block}
\end{figure}

 Due to the complexity of infrared images under different scenes, it is necessary to enable the network to capture basic features of the infrared image for simplifying the small target detection task.
 To avoid omitting small target features that usually take up only a few pixels, all of pooling layers are threw away in the global feature extractor.
 The architecture of the global feature extractor is shown in Fig.~\ref{fig:architecture of GFE and residual block}.
 In this paper, the number $L_g$ of residual blocks used in this module is set to $4$, and the number of convolution kernels for each block is $64$, $128$, $64$ and $8$, respectively.
 While a single-channel infrared image as input $I$ is fed to the global feature extractor, the output data are several global feature maps as the same size as the input $I$.

\subsection{Supervised Attention Module}
\label{subsec:supervised attention module}

\begin{figure}[htbp]
\centerline{\includegraphics[scale=0.56]{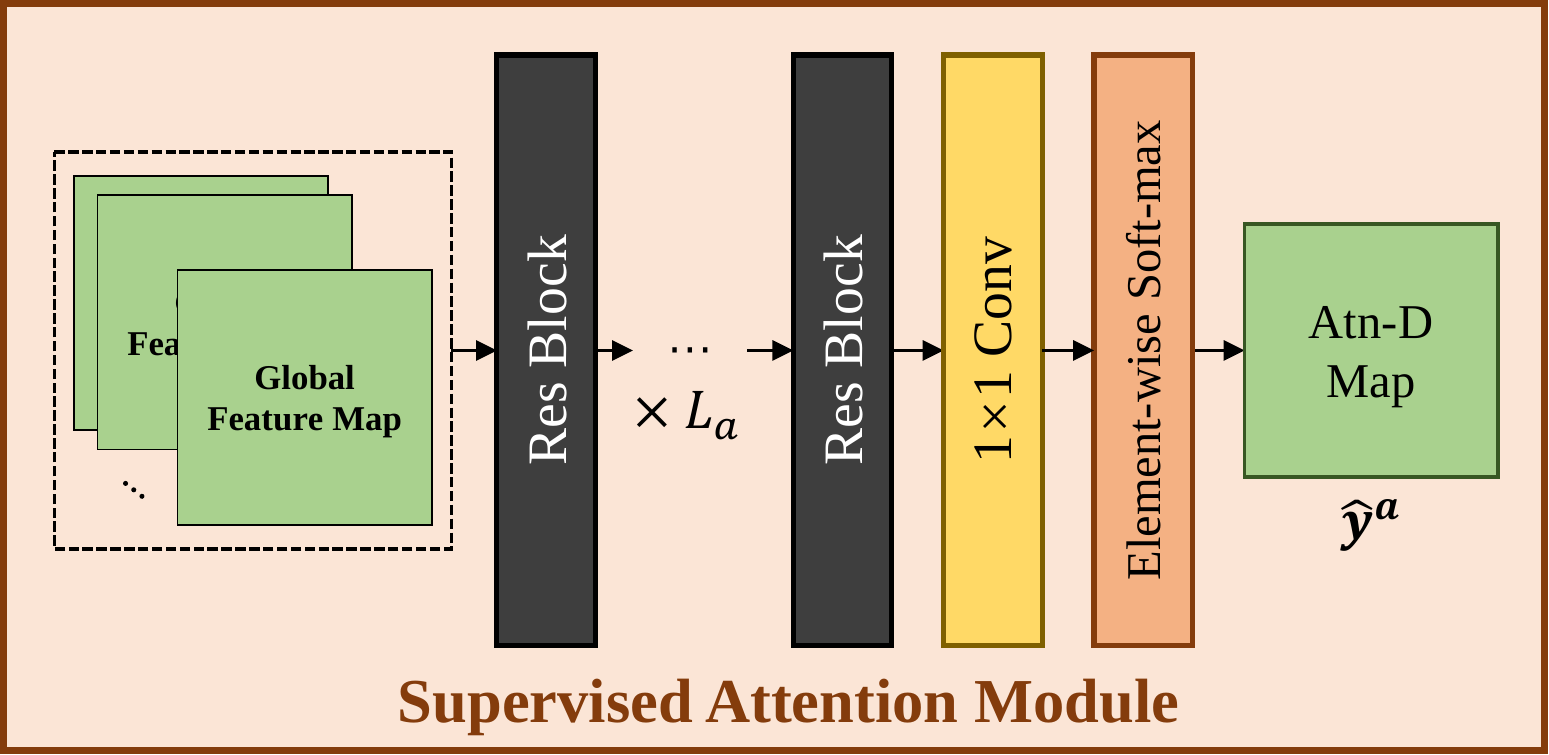}}
\caption{
 The architecture of the supervised attention module.
 The element-wise soft-max layer is a softmax active function applied on all elements to enhance regions of interest.
}
\label{fig:architecture of SAM}
\end{figure}

\begin{figure*}[htbp]
\centerline{\includegraphics[scale=0.624]{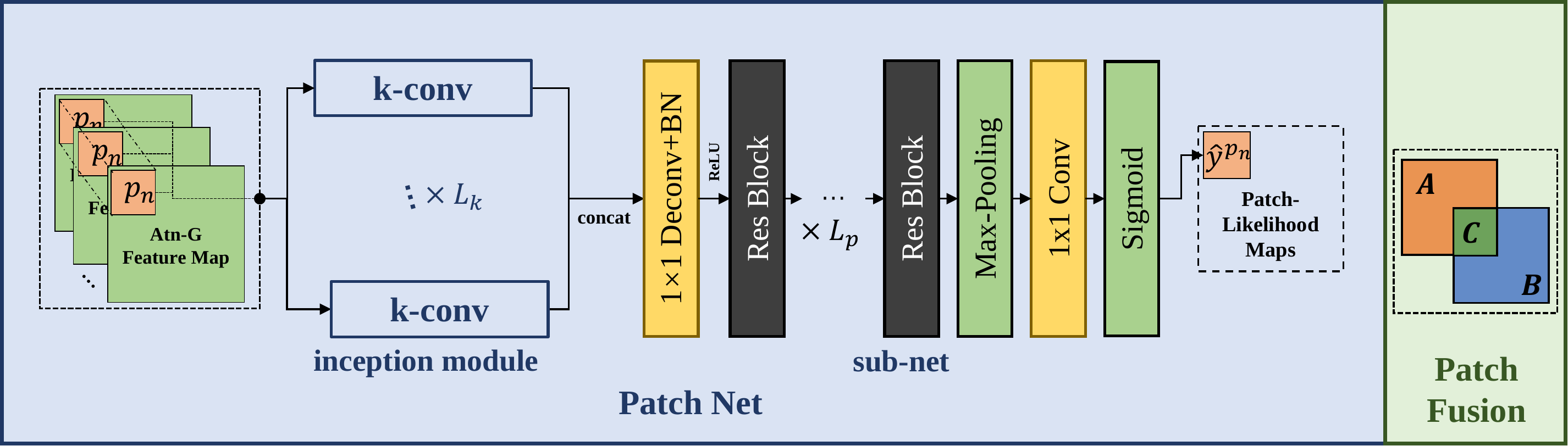}}
\caption{
 The architecture of the patch net and illustration of the patch fusion.
 In the patch net, an inception module composed of several k-conv layers with different kernel size is used for extracting multi-scale features from each local patch $p_n$.
 The sub-net is used for capturing small target features and selecting the patches that contain small targets.
 The patch fusion is used for fusing all patches to get the confidence map.
 In the patch fusion, the value of a pixel in patch $C$ is the mean value of corresponding pixels from all patches (e.g., the patch $A$ and patch $B$ in this figure) covering this pixel.
}
\label{fig:patch net}
\end{figure*}

 For the case of the complicated background such as bright clutters, most conventional methods could not perform well and state-of-the-art methods heavily depended on large amounts of data with heterogeneous scenes.
 In this paper, this issue is addressed through applying an attention module to suppress the background and enhance small targets.
 Through suppressing most irrelevant background pixels, the complexity of extracting small target features from local patches for the patch net can be greatly reduced, which is conductive to train the network on a small dataset with a large number of complex scenes.
 This is very important in practical applications, because it could be very difficult or expensive to collect infrared data for some specific scenes.

 Although the attention module can be trained by the non-supervised way, it will take too much time and iterations to achieve the convergence.
 Thus, we use the supervised way to train the attention module by transformed the ground-truth.
 Due to the sparsity of small targets and the extreme class-imbalance between small targets and the background, it is difficult to directly train the attention module by the ground-truth.
 A feasible way is transforming the ground-truth by a low pass filter into a target spread map which focuses on the adjacent region of small targets.
 By this way, most background suppression and rough target enhancement can be achieved in the attention module, while the precise segmentation will be assigned to the patch-net.
 
 Specifically, we utilize a 2D-Gaussian Low Pass Filter described as:
\begin{small}
\begin{eqnarray}
G=e^{-\frac{1}{2}(\frac{u^{2}+v^{2}}{\sigma^{2}})}, \label{eq:2d-gaussian low pass filter}
\end{eqnarray}
\end{small}where the frequency filtering range is defined by extent parameter $\sigma$ and $u$, $v$ represent the 2-D components in the frequency domain.
 This filter is applied on the ground-truth in the frequency domain and the filtered result $f'$ is transformed back to the spatial domain by Inverse Discrete Fourier Transform (IDFT).
%  This process is described as:
% \begin{small}
% \begin{eqnarray}
% f' = f^{gt} \circledast g =\mathcal{IDFT}\{F^{gt} \times G\}, \label{eq:filtering process}
% \end{eqnarray}
% \end{small}where the $f^{gt}$ and $F^{gt}$ represent the ground-truth image in spatial domain and frequency domain, respectively, while the $g$ and $G$ represent the gaussian filter in spatial domain and frequency domain, respectively.
%  The symbol $\circledast$ is a convolution operator and $f'$ is the filtered result.
 Finally, the target spread map is calculated by:
\begin{small}
\begin{eqnarray}
m_{t} = \frac{f'}{\|f'\|_1}, 
\label{eq:target spread map}
\end{eqnarray}
\end{small}where $m_{t}$ represents the target spread map which indicates some regions that the model should pay attention to and the pixel value of $m_{t}$ represents how much attention that the detector should pay.
 The architecture of the supervised attention module is shown in Fig.~\ref{fig:architecture of SAM}.
 Specifically, the number $L_a$ of residual blocks used in this module is set to $6$, and the number of convolution kernels for each residual block is $32$, $64$, $128$, $64$, $32$ and $1$, respectively.
 The $1\times1$ convolution layer aims to allocate an probability for each element to get an attention-probability map.
 The element-wise soft-max layer is applied on attention-probability map to get the Atn-D map $\hat{y}^{a}$.
 This element-wise soft-max layer is helpful to enhance the regions of interest for accelerating the convergence in training.
 After that, through multiplying the Atn-D map $\hat{y}^{a}$ with each global feature map by element-wise, most background will be suppressed and thus small targets will be enhanced.
 The result of the element-wise multiplication between the Atn-D map $\hat{y}^{a}$ and a global feature map is an Atn-G feature map.

\subsection{Patch Net}
\label{subsec:patch net}
 In the Atn-D map from the supervised attention module, there are many rough target possible regions which are not enough for acquiring the final segmentation due to the high FA rate in the Atn-D map.
 The patch-net is designed to reduce the FA rate and acquire the precise segmentation results by effectively extracting local features correlated with small targets.
 The architecture of the patch net is shown in Fig.~\ref{fig:patch net}.
 In the patch net, each Atn-G feature map is split into a series of patches by using a sliding window.
 There are two circumstances in these patches:
 1) A patch totally belongs to the background.
 Under this circumstance, the patch net is prone to transform this patch to a black block that all pixel values equal to zeros.
 This operation does not demand any computational complexity because these background pixels have been suppressed in the supervised attention module and already similar to zeros.
 2) A patch includes features correlated with small targets.
 Under this circumstance, the patch net will focus on the local saliency of small targets because the sliding window has restricted the receptive field of the patch net on this patch, so that the precise result can be acquired.
 Besides, by sharing the same convolution weights for all patches, the number of parameters of the patch net can also be greatly reduced.

 In this paper, we utilize a sliding window to extract patches from each Atn-G feature map.
 Inspired by~\cite{7298594}, an inception module is used to extract multi-scale small target features from each patch.
 This inception module consists of several k-convs with different kernel sizes and each k-conv is composed of a convolution kernel, a batch norm layer and a ReLU function.
 Specifically, the number $L_k$ of k-convs used in this module is set to $3$, and the convolution kernel size of each k-conv is $1\times1$, $3\times3$ and $5\times5$, respectively.
 Different kernel sizes could be conducive to extract multi-scale features of small targets with various sizes.
 These multi-scale features from each patch are fed to a sub-net for extracting a local feature map named patch-feature map in this paper.
 In the sub-net, each input is firstly scaled to double times through a deconvolution layer which consists of a $1\times1$ deconvolution kernel, a batch norm layer and a ReLU function.
 The deconvolution layer aims to increase the size of small targets in each patch for expanding the perception range of the patch net on small targets.
 Then several residual blocks are used to extract local features from deconvoluted patches.
 The number $L_p$ of these residual blocks used in this module is set to $3$, and the number of convolution kernels in each residual block is $32$, $32$ and $1$, respectively.
 A max pooling layer is utilized to resize each patch-feature map to the original size as the same as the input patch.
 Finally, a single-channel $1\times1$ convolution layer and the sigmoid function are applied to convert each resized feature map to a patch-likelihood map, in which the pixel value represents the probability that the pixel belongs to a small target.

 To get the final confidence map, a patch fusion is proposed to fuse all patch-likelihood maps.
 The patch fusion used in this paper is directly to calculate the mean value of overlapped regions of different patches, as shown in Fig.~\ref{fig:patch net}.
 The pixel values in the region $C$ are the mean value of overlapped pixels from $A$ and $B$.
 Through this strategy, those pixels regarded by most patch-likelihood maps as small target pixels would be enhanced, so that the FA rate will be reduced.
%  To make all fused values in the range of $\left[0, 1\right]$, we adopt a min-max normalization function as follows:
% \begin{small}
% \begin{eqnarray}
% \mathcal{norm}(x) = \frac{x-\min{\{x\}}}{\max{\{x\}}-\min{\{x\}}}.
% \label{eq:min-max function}
% \end{eqnarray}
% \end{small}

 An adaptive threshold $t_{adpt}$~\cite{gao2013infrared} is used to segment the confidence map $\hat{y}$ to get the final prediction result:
\begin{small}
\begin{eqnarray}
t_{adpt} = \max{\{v_{min}, \mu+k\sigma\}},
\label{eq:adaptive threshold}
\end{eqnarray}
\end{small}where $k$ and $v_{min}$ are constants determined experientially.
 $v_{min}$ is to process the case with no target.
 If the max value of the confidence map is lower than $v_{min}$, it means that there is no target in the predicted result.
 $\mu$ and $\sigma$ are the mean value and standard deviation of the confidence map $\hat{y}$.
% defined as follows:
% \begin{small}
% \begin{eqnarray}
% \hat{y}_{pos} = 
% \left\{
% \begin{array}{ll}
%  \hat{y}, & if\quad\hat{y} \geq 0.5, \\
%  \mathcal{nan}, & otherwise,
% \end{array}
% \right.
% \label{eq:positive confidence map}
% \end{eqnarray}
% \end{small}where $\mathcal{nan}$ means that the value less than $0.5$ in $\hat{y}$ will not be involved in the calculation of $\mu_{pos}$ and $\sigma_{pos}$.
 Therefore, a pixel at $(i,j)$ can be segmented as the target pixel if $\hat{y}(i,j) \geq t_{adpt}$, otherwise it is a background pixel.

\subsection{Loss Function}
\label{subsec:loss function}
 The proposed method can be trained in an end-to-end fashion.
 We use the infrared image $I$, its ground-truth image $y$ and its target spread map $m_{t}$ calculated by Eq.~(\ref{eq:target spread map}) to train the proposed method.

 The attention loss is calculated by the sum-square error (SSE) between the target spread map $m_{t}$ and the Atn-D map $\hat{y}^{a}$ of the supervised attention module, defined as follows:
\begin{small}
\begin{eqnarray}
loss_{a} = {\|\hat{y}^{a}-m_t\|}^{2}_{2}. \label{eq:attention loss}
\end{eqnarray}
\end{small}

 The loss of each patch is calculated by the binary cross entropy (BCE) described as follows:
\begin{small}
\begin{eqnarray}
loss_{p_n} = \sum_{i\in{p_n}}{-y_i^{gt_n}\log{\hat{y}_i^{p_n}}-(1-y_i^{gt_n})\log{(1-\hat{y}_i^{p_n})}}, \label{eq:patch loss}
\end{eqnarray}
\end{small}where $y^{gt_n}$ represents the $n$-th patch in the ground-truth image $y$, while $\hat{y}^{p_n}$ represents the corresponding patch-likelihood map.

 The total loss function is defined as:
\begin{small}
\begin{eqnarray}
loss = loss_{a} + \frac{1}{N}\sum^N_{n=1}{loss_{p_n}}, \label{eq:all loss}
\end{eqnarray}
\end{small}where $N$ is the number of patches.

\section{Experiments}
\label{sec:experiments and analysis}
 In this section, we firstly introduce evaluation metrics, state-of-the-art methods used for comparison and experimental settings of all tested methods.
 Then the quantitative comparisons with state-of-the-art methods are performed on two public widely used datasets and one private dataset collected by ourselves, respectively.
 Qualitative results are also given for intuitively illustrating the superiority of our method.
 Finally, the effect of each module, operation and hyperparameter of the proposed method are discussed through a series of ablation studies.

\subsection{Experimental Setup}
\label{subsec:experimental setup}
 
\subsubsection{Evaluation Metrics}
\label{subsubsec:evaluation metrics}
 In early stages, some evaluation metrics widely used in the small target detection community~\cite{gao2013infrared, zhu2020tnlrs, chen2013local} are based on target-level.
 In these metrics, a connected domain in the detection result will be regarded as a true-positive result if (1) the connected domain has overlapped pixels with the ground-truth and (2) the center pixel distance of the connected domain and the ground-truth is within a threshold (usually 4 pixels)~\cite{gao2013infrared}.
%  However, with the development of small target detection technique, the performance of detectors on target-level based metrics becomes better and better, and even could achieve more than $90.00\%$ on probability of detection ($P_d$).
 Additionally, some pixel-level evaluation metrics are also widely used in deep learning based methods~\cite{zhao2019tbc, wang2019miss, shi2019infrared, zhao2020novel}.
 Generally, these pixel-level metrics are more rigorous than target-level metrics when evaluating the performance of a method.
 However, due to the inevitable artificial errors when labeling the ground-truth images, the pixel-level metrics are not enough to provide comprehensive evaluations as well.
 Thus, we use both target-level metrics and pixel-level metrics to comprehensively evaluate all methods.
 
 Specifically, the metrics used for comparison between the proposed method and state-of-the-art methods are as follows:
 \begin{itemize}
     \item The probability of detection ($P_d$) and false alarm rate ($F_a$):
     These two target level metrics are widely used to evaluate the method on infrared small target detection, which are defined as~\cite{gao2013infrared}:
    \begin{small}
    \begin{eqnarray}
    P_d = \frac{\text{\# number of true detections}}{\text{\# number of real targets}}, \label{eq:probability of detection}
    \end{eqnarray}
    \end{small}
    \begin{small}
    \begin{eqnarray}
    F_a = \frac{\text{\# number of false detections}}{\text{\# number of images}}. \label{eq:false alarm rate}
    \end{eqnarray}
    \end{small}
    
     We use $P_d$ with $F_a=0.2/image$ as\cite{gao2013infrared} to evaluate the performance of each method and the area under $P_d$-$F_a$ curve (AUC) with $F_a\leqslant2.0/image$ to evaluate average performance of each method.
     \item Target-level precision, recall and f1 measure ($F1_T$):
     We also use the target-level precision, recall and f1 measure $F1_T$ to evaluate the proposed method and state-of-the-art methods.
     The highest $F1_T$ of each method is provided to display the best performance, and corresponding precision and recall are provided for analysis.
     \item Pixel-level precision, recall and f1 measure ($F1_P$):
     For providing comprehensive comparison, we utilize pixel-level metrics to evaluate each method as well.
     As the same as deep learning methods~\cite{zhao2019tbc, wang2019miss, shi2019infrared, zhao2020novel}, the highest $F1_P$ and corresponding precision and recall are provided.
    %  \item The testing speed ($T_s$):
    %  When evaluating a infrared small target detector, $T_s$ is a pivotal metric which directly determines the application prospect of the tested method.
    %  The faster $T_s$ enables the method to be applied more widely.
    %  We compare the summation of $T_s$ for each method on detecting all images from test set.
 \end{itemize}

\subsubsection{State-of-the-Art Methods}
\label{subsubsec:state-of-the-art methods}
We compare the proposed method with two groups of related methods:
\begin{itemize}
     \item Conventional methods:
     Top-Hat~\cite{deng2018adaptive, bai2010analysis, zeng2006design}, Max-Mean/Max-Median~\cite{deshpande1999max, han2020infrared}, AAGD~\cite{aghaziyarati2019small}, ADMD~\cite{moradi2020fast}, GST~\cite{gao2008generalised},
     LIG~\cite{zhang2018infrared}, ILCM~\cite{han2014robust}, MPCM~\cite{wei2016multiscale}, TLLCM~\cite{8922738}, LEF~\cite{xia2019infrared},
     IPI~\cite{gao2013infrared}.
     \item Deep learning method:
         MDvsFA-cGAN~\cite{wang2019miss}, ACM~\cite{dai21acm}.
        %  \textcolor{red}{Although some deep learning methods are also proposed in recent years, the source code of these methods is rarely open for public.}
         These two methods achieve excellent performance to the best of our knowledge so that are used to represent the state-of-the-art deep learning methods.
\end{itemize}
 
\subsubsection{Experimental Setting}
\label{subsubsec:experimental setting}
 The parameter settings for all tested methods are listed in Tab.~\ref{tab:parameter setting}.
 The experiment is conducted on a computer with one 3.40GHz CPU, 32GB RAM and two NVIDIA TITAN V GPUs.
 All trainable methods are trained from the scratch, and the batch size is set to $20$ for training.
 The proposed method is implemented by Python and PyTorch.
 Specifically, the input image is resized to $120\times120$, the patch size and sliding step are set to $30\times30$ and $10$, respectively.
 Adam algorithm~\cite{kingma2014adam} is used as optimizer and learning rate is set to $0.001$ for all batches.

\begin{table}[t]
\caption{Parameter setting for tested methods.}
% \vspace{-10pt}
\label{tab:parameter setting}
\begin{center}
\setlength{\tabcolsep}{2mm}{
\begin{tabular}{lc}
\hline
\hline
Methods & Key parameter settings \\
\hline
Top-Hat~\cite{zeng2006design} & structure size: $12\times12$ \\[3pt]

Max-Mean/Max-Median~\cite{deshpande1999max} & filter size: $15\times15$ \\[3pt]

AAGD~\cite{aghaziyarati2019small} & scale size: 3, 5, 7, 9 \\[3pt]

ADMD~\cite{moradi2020fast} & scale size: 3, 5, 7, 9 \\[3pt]

LIG~\cite{zhang2018infrared} & $k=0.2$, $N=11$ \\[3pt]

IPI~\cite{gao2013infrared} & patch size: $50\times50$, sliding step: $10$, \\ & $\lambda=(1/\sqrt{\max{(m,n)}})$, $\epsilon=10^{-6}$ \\[3pt]

ILCM~\cite{han2014robust} & subblock size: $8\times8$, moving step: $4$ \\[3pt]

MPCM~\cite{wei2016multiscale} & scale size: 3, 5, 7 \\[3pt]

TLLCM~\cite{8922738} & gaussian kernel size: $3\times3$, \\ & scale size: 3, 5, 7, 9 \\[3pt]

LEF~\cite{xia2019infrared} & $\alpha=0.5$, $h=0.2$, \\ & scale size: 3, 5, 7, 9 \\[3pt]

GST~\cite{gao2008generalised} & $\sigma_{1}=0.6$, $\sigma_{2}=1.1$, \\ & boundary width: $5$, filter size: $5\times5$ \\[3pt]

MDvsFA-cGAN~\cite{wang2019miss} & image size: $128\times128$, \\ & $\lambda_1=100$, $\lambda_2=10$, \\ & trainable parameters: 3919598 \\[3pt]

ACM~\cite{dai21acm} & image size: $256\times256$, \\ & backbone: fpn, fuse: asymbi, \\ & trainable parameters: 387187 \\[3pt]

Ours & image size: $120\times120$, \\ & patch size: $30\times30$, sliding step: $10$, \\ & trainable parameters: 925108 \\
\hline
\hline
\end{tabular}
}
\end{center}
\end{table}

\subsection{Quantitative Analysis}
\label{subsec:quantitative analysis}

 To evaluate the performance of the proposed method, we use three datasets with differences, including a synthetic dataset namely MFIRST used in~\cite{wang2019miss}, a widely used SIRST dataset~\cite{dai21acm} and a SeqIRST dataset which consists of seven infrared sequences collected by ourselves.
 
\begin{figure}[h]
	{\scriptsize{(a)}}
	\subfigure{
		\begin{minipage}[htp]{0.45\textwidth}
			\centering
			\includegraphics[width=1\textwidth]{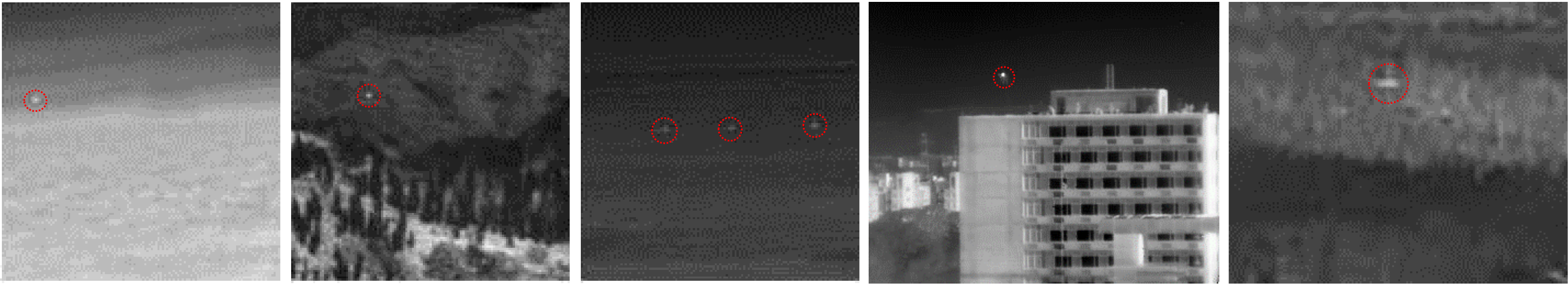}
		\end{minipage}
	}\vspace{-1mm}

	{\scriptsize{(b)}}
	\subfigure{
		\begin{minipage}[htp]{0.45\textwidth}
			\centering
			\includegraphics[width=1\textwidth]{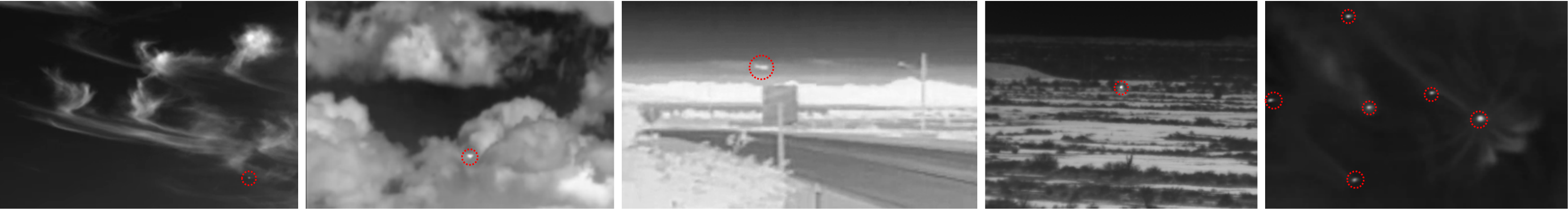}
		\end{minipage}
	}\vspace{-1mm}

	{\scriptsize{(c)}}
	\subfigure{
		\begin{minipage}[htp]{0.45\textwidth}
			\centering
			\includegraphics[width=1\textwidth]{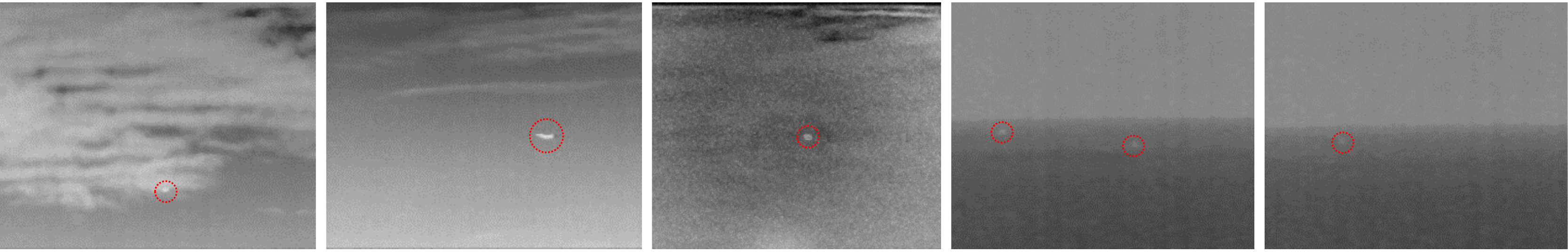}
		\end{minipage}
	}\vspace{-1mm}
	\caption{Representative images from (a) MFIRST, (b) SIRST and (c) SeqIRST.
	Small targets to segment are indicated by red circles.
	Single-frame infrared images in the SIRST and infrared sequences in the SeqIRST are collected under real scenes, while single-frame infrared images are synthesized through coding in the MFIRST.
	}
	\label{fig:representative images of three datasets}
\end{figure}

 Some representative images of these three datasets are shown in Fig.~\ref{fig:representative images of three datasets}, while the training and testing splittings of these three datasets are listed in Tab.~\ref{tab:training and testing splittings}.
 
\begin{table}[t]
\caption{Training/testing splitting and datasets illustration.}
\label{tab:training and testing splittings}
\begin{center}
\setlength{\tabcolsep}{2mm}{
\begin{threeparttable}
\begin{tabular}{lcccc}
\hline
\hline
Datasets & Training Images & Testing Images & Target Simulation \\ \hline
MFIRST  & 9960 & 100 & + \\
SIRST & 341 & 86 & - \\
SeqIRST & 465 & 75 & - \\
\hline
\hline

\end{tabular}
\begin{tablenotes}
    \footnotesize{
    \item[] `+' / `-' means that the target simulation technique is / is not applied on the dataset, respectively.
}
\end{tablenotes}
\end{threeparttable}
}
\end{center}
\end{table}

\subsubsection{MFIRST dataset}
\label{subsubsec:mfirst dataset}
 The MFIRST is a dataset containing a large number of real and synthetic infrared images.
 The real infrared images come from two bespoke datasets containing 11 real infrared sequences with 2098 frames in total and 100 real individual infrared images with different small objects, respectively~\cite{wang2019miss}.
 In the synthetic infrared images, the backgrounds are generated by cropping different regions from infrared high-resolution natural scene images and the small targets are separated from the real infrared images or synthesized by using the 2-D Gaussian function~\cite{wang2019miss}.
 We use this dataset to evaluate the performance of each method on augmented data with a large number of different scenes.
 The comparison results on the MFIRST dataset are listed in Tab.~\ref{tab:comparison on mfirst}.

\begin{table}[t]
\caption{Comparison on the MFIRST dataset.}
\label{tab:comparison on mfirst}
\begin{center}
\setlength{\tabcolsep}{1.2mm}{
\begin{threeparttable}
\begin{tabular}{l|cccc}
\hline
\hline
Methods & \multicolumn{4}{c}{MFIRST} \\

& \multicolumn{1}{c}{$P_d$ (\%)} & \multicolumn{1}{c}{AUC (\%)} & \multicolumn{1}{c}{$F1_T$ (\%)} & \multicolumn{1}{c}{$F1_P$ (\%)} \\
\hline

Top-Hat~\cite{zeng2006design} & - & - & 44.62 & 12.80 \\

Max-Mean/Max-Median~\cite{deshpande1999max} & - & 50.50 & 58.30 & 14.44 \\

AAGD~\cite{aghaziyarati2019small} & 43.66 & 56.96 & 65.70 & 32.42 \\

ADMD~\cite{moradi2020fast} & 59.64 & 64.09 & 70.99 & 31.52 \\

LIG~\cite{zhang2018infrared} & 59.29 & 64.17 & 70.87 & 41.27 \\

IPI~\cite{gao2013infrared} & 41.59 & 51.02 & 60.73 & 33.58 \\

ILCM~\cite{han2014robust} & - & - & 24.52 & 0.91 \\

MPCM~\cite{wei2016multiscale} & 57.86 & 64.62 & 72.20 & 35.43 \\

TLLCM~\cite{8922738} & - & 46.43 & 52.63 & 6.67 \\

LEF~\cite{xia2019infrared} & 49.49 & 70.01 & 72.45 & 5.87 \\

GST~\cite{gao2008generalised} & 56.39 & 59.69 & 66.67 & 24.67 \\

MDvsFA-cGAN~\cite{wang2019miss}& 78.16 & 79.51 & 86.31 & 60.03 \\

ACM~\cite{dai21acm} & 74.89 & 75.52 & 85.61 & 56.10 \\

% \textcolor{red}{ACM} & \textcolor{red}{70.07} & \textcolor{red}{71.95} & \textcolor{red}{82.11} & \textcolor{red}{58.05} \\

\textbf{Ours} & \textbf{79.29} & \textbf{81.14} & \textbf{87.32} & \textbf{60.07} \\
\hline
\hline
\end{tabular}
\begin{tablenotes}
    \footnotesize{
    \item[] `-' means that the method can not get reasonable values under fixed $F_a=0.2/image$ for $P_d$ or under $F_a\leqslant2.0/image$ for AUC.
    }
\end{tablenotes}
\end{threeparttable}
}
\end{center}
\end{table}

 On the MFIRST dataset, due to the large amount of data, it is hard for prior based conventional methods to be generalized enough for discrepant scenes so that these methods perform much worse than data-driven methods.
 Compared with the deep learning method, the proposed method outperforms MDvsFA-cGAN approximately $2\%$ on $AUC$, and outperforms ACM approximately $4\%$ on both metrics.
 Although the $P_d$, $F1_T$ and $F1_P$ metrics of the proposed method are similar with MDvsFA-cGAN, the higher $AUC$ illustrates the better average performance of the proposed method.
 By the attention mechanism and the patch splitting, the proposed method addresses the class-imbalance of small target pixels and background pixels.
 Focusing on the local saliency of small targets enables the proposed method to capture more effective features of small targets for achieving robust detection results.

% {\color{red}{
%  On the MFIRST, prior based conventional methods and deep learning methods both have favorable performance.
%  Compared with conventional methods, deep learning methods have much better $F1_P$ because these data driven methods could learn some boundary and shape characteristics (even so weak) of small targets.
%  By attention mechanism and patch splitting, the proposed method addresses the class-imbalance of small target pixels and background pixels and focuses on local saliency of small targets, so that could capture more effective features of small targets.
%  Compared with MDvsFA-cGAN, the best state-of-the-art method, the proposed method has similar $F1_P$ and outperforms approximately $1\%$ on $P_d$, $2\%$ on $AUC$ and $1\%$ on $F1_T$, which demonstrates the superiority of LPNet on augmented data with complicated background.
%  }}

\subsubsection{SIRST dataset}
\label{subsubsec:sirst dataset}

 The SIRST dataset is a widely used public dataset for single-frame infrared small target detection~\cite{wang2019miss}.
 It contains 427 representative images and 480 instances of different scenes from hundreds of real-world videos.
 We use this dataset to evaluate the performance of each method on limited but authentic data with a large number of different real scenes.
 The comparison results on the SIRST dataset are listed in Tab.~\ref{tab:comparison on sirst}.

\begin{table}[t]
\caption{Comparison on the SIRST dataset.}
\label{tab:comparison on sirst}
\begin{center}
\setlength{\tabcolsep}{1.2mm}{
\begin{threeparttable}
\begin{tabular}{l|cccc}
\hline
\hline
Methods & \multicolumn{4}{c}{SIRST} \\

& \multicolumn{1}{c}{$P_d$ (\%)} & \multicolumn{1}{c}{AUC (\%)} & \multicolumn{1}{c}{$F1_T$ (\%)} & \multicolumn{1}{c}{$F1_P$ (\%)} \\

\hline

Top-Hat~\cite{zeng2006design} & 85.34 & 82.38 & 82.52 & 44.13 \\

Max-Mean/Max-Median~\cite{deshpande1999max} & 78.46 & 77.45 & 73.49 & 23.97 \\

AAGD~\cite{aghaziyarati2019small} & 89.09 & 88.14 & 84.69 & 50.27 \\

ADMD~\cite{moradi2020fast} & 94.13 & 90.46 & 88.50 & 56.69 \\

LIG~\cite{zhang2018infrared} & 90.19 & 90.00 & 89.72 & 59.15 \\

IPI~\cite{gao2013infrared} & 86.87 & 84.45 & 85.32 & 56.97 \\

ILCM~\cite{han2014robust} & - & - & 47.26 & 0.71 \\

MPCM~\cite{wei2016multiscale} & 93.56 & 90.40 & 86.96 & 58.59 \\

TLLCM~\cite{8922738} & 61.61 & 79.14 & 79.66 & 7.60 \\

LEF~\cite{xia2019infrared} & - & - & 57.60 & 2.45 \\

GST~\cite{gao2008generalised} & 77.01 & 76.81 & 80.40 & 35.32 \\

% MDvsFA-cGAN~\cite{wang2019miss} & - & - & - & - \\

ACM~\cite{dai21acm} & 93.67 & 86.93 & 95.85 & 69.61 \\

% \textcolor{red}{ACM} & \textcolor{red}{98.24} & \textcolor{red}{91.67} & \textcolor{red}{96.77} & \textcolor{red}{81.30} \\

\textbf{Ours} & \textbf{97.40} & \textbf{97.86} & \textbf{96.40} & \textbf{71.66} \\
\hline
\hline
\end{tabular}
\begin{tablenotes}
    \footnotesize{
    \item[] `-' means that the method can not get reasonable values under fixed $F_a=0.2/image$ for $P_d$ or under $F_a\leqslant2.0/image$ for AUC.
    \item[] MDvsFA-cGAN~\cite{wang2019miss} does not get reasonable values in our experiment due to the failure of convergence on such small dataset.
    }
\end{tablenotes}
\end{threeparttable}
}
\end{center}
\end{table}

 On the SIRST dataset, prior based conventional methods perform well because the amount of data in the SIRST is limited and it is easy for these methods to cover all complicated scenes by some specific expert knowledge.
 Even so, deep learning methods still have better performance on both metrics when compared with prior based conventional methods.
 Compared with other deep learning methods and conventional methods, the proposed method achieves the best performance on all metrics (approximately promoting $3\%$ on $P_d$, $7\%$ on AUC and $F1_T$, and $3\%$ on $F1_P$).
 The state-of-the-art performance of the proposed method demonstrates the effectiveness of the attention mechanism and the patch splitting on effectively extracting small target features when being trained on limited data for deep learning methods, as described in Section~\ref{section:introduction}.

% {\color{red}{
%  Among state-of-the-art methods, 
 
%  conventional methods perform better than deep learning methods because most deep learning methods rely on numerous data to extract effective small target features, especially with complicated scenes.
%  When the SIRST dataset is limited and complicated, the representative deep learning method, MDvsFA-cGAN, fails to adaptively capture effective features and thus is worse than prior based conventional methods.
%  However, through introducing supervised attention module to suppress most irrelevant background pixels, the complexity of background is extremely reduced so that the proposed method possesses the ability of extracting effective small target features from limited data with heterogeneous scenes, as described in Sec.~\ref{subsec:supervised attention module}.
%  Specifically, the proposed method outperforms the best state-of-the-art method approximately $6\%$ on target-level metrics and $10\%$ on pixel-level metrics, which demonstrates the superiority of LPNet on limited data with heterogeneous scenes.
% }}

\subsubsection{SeqIRST dataset}
\label{subsubsec:sirst dataset}

\begin{table}[t]
\caption{Details of the SeqIRST dataset.}
\label{tab:details of SeqIRST}
\begin{center}
\setlength{\tabcolsep}{6mm}{
\begin{tabular}{lcccc}
\hline
\hline
No. & Target  & Background & Frames \\
\hline
Seq. 1 & plane & sky (cloudy) & 30 \\
Seq. 2 & plane & sky (cloudy) & 39 \\
Seq. 3 & plane & sky & 39  \\
Seq. 4 & bird & sky (cloudy) & 123 \\
Seq. 5 & vessel & sea & 100 \\
Seq. 6 & plane & sky & 109 \\
Seq. 7 & vessel & sea & 100 \\
\hline
\hline
\end{tabular}
}
\end{center}
% \vspace{-10pt}
\end{table}

 The SeqIRST dataset contains seven infrared sequences which are collected by ourselves and the details are listed in Tab.~\ref{tab:details of SeqIRST}.
 It contains three types of small targets under two scenes.
 We use this dataset to evaluate the performance of each method on detecting infrared small targets from continuous frames.
 The comparison results are listed in Tab.~\ref{tab:comparison on seqirst}.

\begin{table}[t]
\caption{Comparison on SeqIRST dataset.}
\label{tab:comparison on seqirst}
\begin{center}
\setlength{\tabcolsep}{1.2mm}{
\begin{threeparttable}
\begin{tabular}{l|cccc}
\hline
\hline
Methods & \multicolumn{4}{c}{SeqIRST} \\

& \multicolumn{1}{c}{$P_d$ (\%)} & \multicolumn{1}{c}{AUC (\%)} & \multicolumn{1}{c}{$F1_T$ (\%)} & \multicolumn{1}{c}{$F1_P$ (\%)} \\

\hline

Top-Hat~\cite{zeng2006design} & - & 58.90 & 61.20 & 10.35 \\

Max-Mean/Max-Median~\cite{deshpande1999max} & - & - & 23.95 & 5.33 \\

AAGD~\cite{aghaziyarati2019small} & - & 56.51 & 60.47 & 17.72 \\

ADMD~\cite{moradi2020fast} & 47.19 & 67.08 & 69.01 & 17.27 \\

LIG~\cite{zhang2018infrared} & - & 57.59 & 60.67 & 20.99 \\

IPI~\cite{gao2013infrared} & 76.42 & 81.54 & 78.45 & 30.95 \\

ILCM~\cite{han2014robust} & - & - & 24.72 & 1.05 \\

MPCM~\cite{wei2016multiscale} & 74.43 & 77.49 & 78.89 & 17.04 \\

TLLCM~\cite{8922738} & 60.44 & 63.33 & 66.26 & 10.67 \\

LEF~\cite{xia2019infrared} & - & 55.68 & 56.65 & 11.23 \\

GST~\cite{gao2008generalised} & - & - & 47.06 & 14.06 \\

MDvsFA-cGAN~\cite{wang2019miss} & 96.33 & 57.00 & 98.88 & 89.68 \\

ACM~\cite{dai21acm} & 96.32 & 84.51 & 99.44 & 89.55 \\

% \textcolor{red}{ACM} & \textcolor{red}{13.09} & \textcolor{red}{42.05} & \textcolor{red}{70.30} & \textcolor{red}{45.20} \\

\textbf{Ours} & \textbf{97.03} & \textbf{98.50} & \textbf{99.99} & \textbf{89.72} \\
\hline
\hline
\end{tabular}
\begin{tablenotes}
    \footnotesize{
    \item[] `-' means that the method can not get reasonable values under fixed $F_a=0.2/image$ for $P_d$ or under $F_a\leqslant2.0/image$ for AUC.
    }
\end{tablenotes}
\end{threeparttable}
}
\end{center}
\end{table}
 
 On the SeqIRST dataset with small amount of data, prior based conventional methods also perform much worse than no-prior methods.
 The reason might be that when small targets in a sequence are too dim, a conventional method will fail to detect targets in all frames if it fails in a single frame.
 Compared with these conventional methods, even if no-prior methods fail to detect a small target in one frame, they could still detect it in other frames since decoupling with the prior correlated with the specific scene.
 Compared with the deep learning method, the proposed method outperforms MDvsFA-cGAN approximately $40\%$ on $AUC$, and outperforms ACM approximately $14\%$ on $AUC$.
 These similar $P_d$, $F1_T$, $F1_P$ and much better $AUC$ on the SeqIRST demonstrate the superiority of the proposed method on average performance again.

% {\color{red}{
%  Compared with prior based conventional methods, deep learning methods have much better performance on SeqIRST dataset.
%  This could be distributed to the reason that continuous frames always have similar backgrounds which reduces the difficulties of deep learning methods to capture effective features even from limited data.
%  By introducing local patch splitting, the proposed method performs better on capturing local features of small targets, which is conductive for detecting small targets with local saliency.
%  Compared with MDvsFA-cGAN that the best state-of-the-art method, LPNet has the similar performance on $F1_P$ and outperforms approximately $1\%$ on $P_d$, $40\%$ on $AUC$ and $2\%$ on $F1_T$.
%  The $F1_T$ and $F1_P$ results illustrate that under the same average pixel-level accuracy, the proposed method has better performance than MDvsFA-cGAN on capturing all small targets.
%  The $P_d$ and $AUC$ results illustrate that the proposed method not only performs better on the highest detection precision but also has much better stability than MDvsFA-cGAN.
% }}

\subsection{Qualitative Analysis}
\label{subsec:qualitative analysis}

\begin{figure*}[htbp]
\centerline{\includegraphics[scale=0.978]{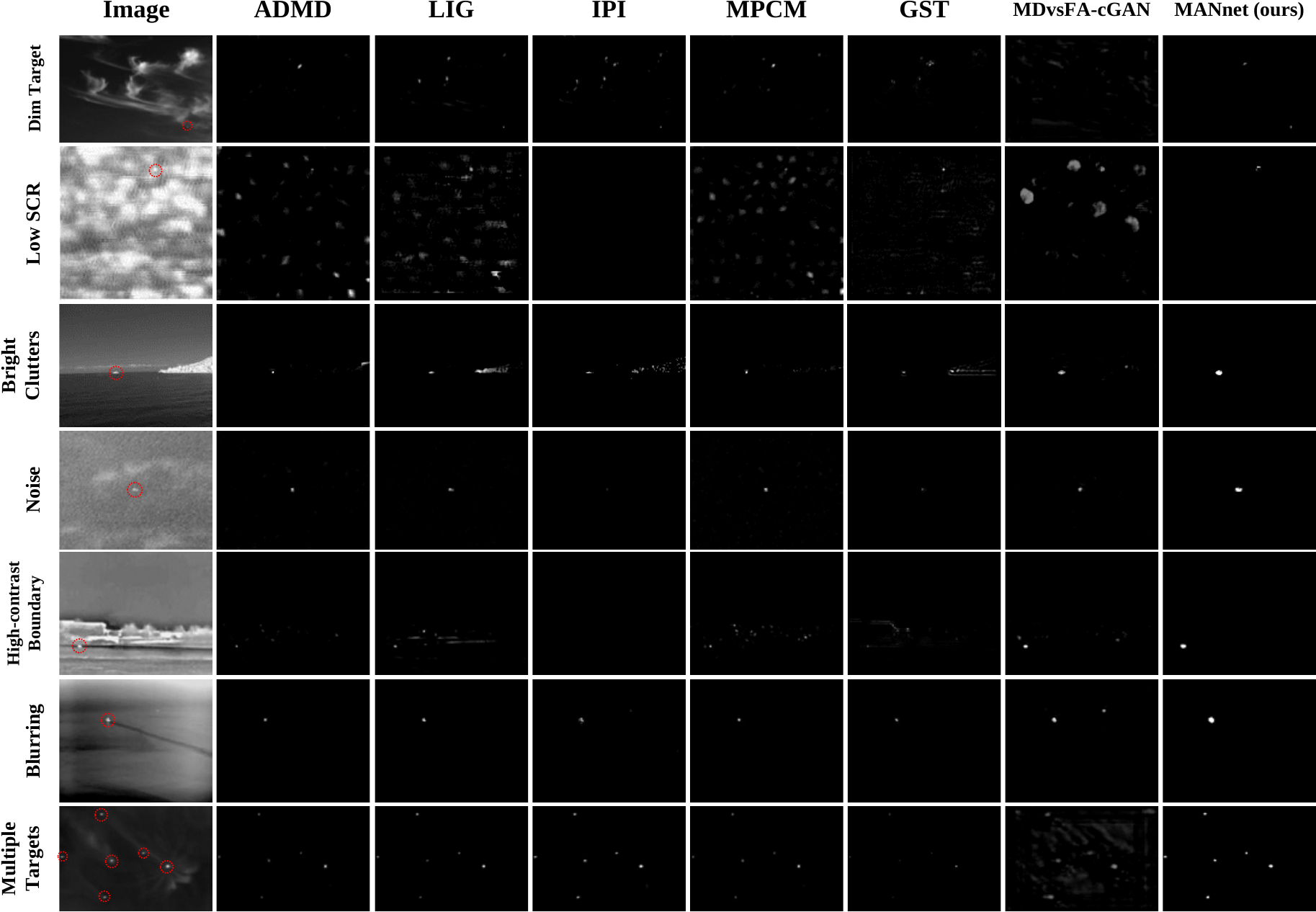}}
\caption{
 The representative processed results of different methods.
 The first row shows results of detecting a dim small target.
 The second row shows results of detecting a small target with low signal-to-clutter ratio (SCR), i.e. the small target buried in clutters of the background.
 The third row shows results of detecting a small target from a image with bright clutters.
 Compared with the second row, the small target in the third row is not buried in clutters but these clutters consist of a series of discrete areas with bright pixels.
 The fourth row shows results of detecting the small target image with high noise level.
 The fifth row shows results of detecting the small target image with high background boundary contrast.
 The sixth row shows results of detecting the small target image with gaussian blurring.
 The last row is results of detecting the small target image with multiple targets.
}
\label{fig:visualization of likelihood maps from six methods}
\end{figure*}

 We give some representative processed results of the proposed method and six methods with top performance, as shown in Fig.~\ref{fig:visualization of likelihood maps from six methods}.
 
 It can be seen from Fig.~\ref{fig:visualization of likelihood maps from six methods} that:
 (1) When detecting the dim target at the first row, most existing methods fail to detect the target and erroneously regard several cloudy clutters of the background as targets.
 (2) When detecting the small target with low signal-to-clutter ratio (SCR), most tested methods have a large number of FA and IPI module fails to detect the target.
 (3) When detecting the small target from the infrared image with bright clutters, most existing methods are interfered by the bright clutters and have a lot of FA in the clutter regions.
 Under above three mainly representative obstacles of infrared small target detection, the proposed method has has the least FA and MD.
 Besides, when facing to other general obstacles of the object detection, such as the high noise level at the fourth row, high boundary contrast of background at the fifth row, gaussian blurring at the sixth row and detecting multiple targets at the last row, the proposed method also possesses detection results with the least FA and MD among tested methods.

\subsection{Ablation Study}
\label{subsec:ablation study}

\subsubsection{Module Discussion}
\label{subsubsec:module discussion}
 To investigate the effect of each modul in the proposed method, we conduct an ablation study on the MFIRST dataset.
 This MFIRST dataset is the largest dataset with numerous heterogeneous scenes among three datasets, which is crucial for validating the comprehensive performance of each ablation block.
 As listed in Tab.~\ref{tab:module discussion}, the AttnNet consists of the global feature extractor, the supervised attention module and a single convolution kernel used to allocate probability for each pixel.
 It is trained by the target spread map with the SSE loss.
 The PatchNet consists of the global feature extractor and the patch net, which is trained by the ground-truth with the BCE loss.

 According to the definition, precision is negatively correlated with FA and recall is negatively correlated with MD, which means that the higher precision represents the lower FA and the higher recall represents the lower MD.
 In AttnNet, the precision is higher than recall, which demonstrates that the attention module has better performance on reducing FA.
 In PatchNet, the recall is higher than precision, which demonstrates that the patch net module has better performance on reducing MD.
 Compared with the AttnNet and PatchNet, our method has both highest precision and recall, which demonstrates that the combination of the attention mechanism and the patch net module is conductive for reducing both FA and MD.
 Besides, the increment of recall from the AttnNet (without the patch net module) to our method is much more than precision, which illustrates that the patch net module has more positive effects on capturing those small targets that might be omitted.
 The increment of precision from the PatchNet (without the attention module) to our method is much more than recall, which illustrates that the attention module has more positive effects on ensuring the final accuracy of detecting.

\begin{table}[t]
\caption{Module Discussion on MFIRST dataset.}
\label{tab:module discussion}
\begin{center}
\setlength{\tabcolsep}{1mm}{
\begin{threeparttable}
\begin{tabular}{l|ccc|ccc}
\hline
\hline
Methods & \multicolumn{3}{c|}{Target Level} & \multicolumn{3}{c}{Pixel Level} \\

& \multicolumn{1}{c}{Prec. (\%)} & \multicolumn{1}{c}{Rec. (\%)} &
\multicolumn{1}{c|}{F1 (\%)}

& \multicolumn{1}{c}{Prec. (\%)} & \multicolumn{1}{c}{Rec. (\%)} & \multicolumn{1}{c}{F1 (\%)} \\
\hline

% CNN\textsuperscript{1} & 58.65 & 60.00 & 59.32 & 37.68 & 43.23 &  40.26 \\
% PatchNet\textsuperscript{2} & 2.74 & 85.71 & 5.31 & 2.56 & 64.88 & 4.92 \\
AttnNet\textsuperscript{1} & 62.96 & 48.57 & 54.84 & 39.40 & 38.89 & 39.14 \\
PatchNet\textsuperscript{2} & 38.25 & 78.42 & 51.42 & 31.27 & 46.27 & 37.32 \\
Ours\textsuperscript{3} & \textbf{86.11} & \textbf{88.57} & \textbf{87.32} & \textbf{52.06} & \textbf{71.00} & \textbf{60.07} \\

\hline
\hline
\end{tabular}
\begin{tablenotes}
    \footnotesize{
    \item[1] AttnNet combines a global feature extractor, a supervised attention module and a single convolution kernel for allocating probability for each pixel.
    The supervised attention module is trained by the target spread map calculated by Eq.~(\ref{eq:target spread map}) with the SSE loss.
    \item[2] PatchNet contains a global feature extractor and a patch net.
    It is trained by the ground-truth with the BCE loss.
    \item[3] Our method is trained by the ground-truth with the BCE loss and the target spread map with the SSE loss jointly.
    }
\end{tablenotes}
\end{threeparttable}
}
\end{center}
\end{table}

\subsubsection{Hyperparameter Discussion}
\label{subsubsec:hyperparameter discussion}
 Besides, we conduct a hyperparameter experiment to illustrate the impact of the patch size and the sliding step on the proposed method, as listed in Tab.~\ref{tab:hyperparameters discussion}.
 With the increasing of the patch size and the sliding step, the performance of the proposed method will be worse and worse because the local property of small targets will lost under a larger and larger patch size, and the patch fusion will also lose effect under a too large sliding step.
 However, a too small patch size also causes the performance degradation.
 This might be because the inception module of the patch net could not capture enough multi-scale features on patches with too small size.

\begin{table}[t]
\caption{Hyperparameters Discussion on MFIRST dataset.}
\label{tab:hyperparameters discussion}
\begin{center}
\setlength{\tabcolsep}{0.76mm}{
\begin{tabular}{cc|ccc|ccc}
\hline
\hline
\multicolumn{2}{l|}{Hyperparameters}
& \multicolumn{3}{c|}{Target Level} & \multicolumn{3}{c}{Pixel Level} \\ 

\multicolumn{1}{c}{Patch} & \multicolumn{1}{c|}{Step} 

& \multicolumn{1}{c}{Prec. (\%)} & \multicolumn{1}{c}{Rec. (\%)} &
\multicolumn{1}{c|}{F1 (\%)}

& \multicolumn{1}{c}{Prec. (\%)} & \multicolumn{1}{c}{Rec. (\%)} & \multicolumn{1}{c}{F1 (\%)} \\
\hline

$20\times20$ & $10$ & 85.93 & 82.86 & 84.36 & 42.81 & 60.72 & 50.22 \\

$30\times30$ & $10$ & \textbf{86.11} & \textbf{88.57} & \textbf{87.32} & \textbf{52.06} & \textbf{71.00} & \textbf{60.07} \\

$30\times30$ & $15$ & 84.92 & 78.10 & 81.37 & 45.98 & 64.59 & 53.72 \\

$40\times40$ & $20$ & 72.80 & 66.91 & 69.73 & 37.45 & 54.89 & 44.52 \\
\hline
\hline
\end{tabular}
}
\end{center}
\end{table}

\subsubsection{Consumption Discussion}
\label{subsubsec:consumption discussion}
 A time and memory consumption experiment is also conducted for evaluating the practicability of each method, as shown in Tab.~\ref{tab:consumption discussion}.
 
\begin{table}[t]
\caption{Consumption Discussion.}
\label{tab:consumption discussion}
\begin{center}
\setlength{\tabcolsep}{3mm}{
\begin{threeparttable}
\begin{tabular}{l|c|c}
\hline
\hline

Methods & Time & GPU Memory \\
        & (s$/$100 images) & (MB) \\
\hline

Top-Hat~\cite{zeng2006design} & 1.78 & - \\

Max-Mean/Max-Median~\cite{deshpande1999max} & \textbf{1.50} & - \\

AAGD~\cite{aghaziyarati2019small} & 3.52 & - \\

ADMD~\cite{moradi2020fast} & 2.02 & - \\

LIG~\cite{zhang2018infrared} & 70.44 & - \\

IPI~\cite{gao2013infrared} & 424.60 & - \\

ILCM~\cite{han2014robust} & 1.92 & - \\

MPCM~\cite{wei2016multiscale} & 4.60 & - \\

TLLCM~\cite{8922738} & 321.91 & - \\

LEF~\cite{xia2019infrared} & 430.22 & - \\

GST~\cite{gao2008generalised} & \textbf{1.05} & - \\

MDvsFA-cGAN~\cite{wang2019miss} & 10.62 & 432.98 \\

ACM~\cite{dai21acm} & 1.61 & \textbf{82.15} \\

\textbf{Ours} & 37.43 & \textbf{363.11} \\

\hline
\hline
\end{tabular}
\begin{tablenotes}
    \footnotesize{
    \item[] `-' means that the method does not possess trainable parameters. The smallest and the second smallest values are shown in boldface.
    }
\end{tablenotes}
\end{threeparttable}
}
\end{center}
\end{table}

 Although the time consumption of the proposed method does not perform best among all tested methods, it could be further reduced by parallelizing the patch splitting that the most time consuming module of the proposed method.
 However, it can be seen from Tab.~\ref{tab:consumption discussion} that the proposed method still has a receivable time and memory consumption compared with most state-of-the-art methods, which ensures the practicability.

\section{Conclusion}
\label{sec:conclusion}
 A new deep learning framework decoupled with prior for infrared small target detection is presented in this paper.
 The global feature extractor enables the proposed method to capture basic features by viewing the whole image.
 The supervised attention module enables the network to suppress background irrelevant with small target features for effectively simplify the task for subsequent modules.
 The patch net enables the network to focus on the local saliency of small target features based on decomposed local patches.
 Extensive synthetic and real data experiments show that compared with state-of-the-art methods, the proposed method performs better on detecting infrared small targets with complicated scenes, such as dim targets, bright clutters and low SCR.
 Besides, the proposed method has higher robustness and comprehensive performance than the compared deep learning method.
 In the future, we will generalize the proposed method from single-frame detection to multi-frame detection by synthesizing spatial and temporal features of infrared small target videos.

\bibliographystyle{unsrt}
\bibliography{main.bib}

\begin{thebibliography}{10}

\bibitem{bai2010analysis}
Xiangzhi Bai and Fugen Zhou.
\newblock Analysis of new top-hat transformation and the application for
  infrared dim small target detection.
\newblock {\em Pattern Recognition}, 43(6):2145--2156, 2010.

\bibitem{zeng2006design}
Ming Zeng, Jianxun Li, and Zhang Peng.
\newblock The design of top-hat morphological filter and application to
  infrared target detection.
\newblock {\em Infrared Physics \& Technology}, 48(1):67--76, 2006.

\bibitem{aghaziyarati2019small}
Saeid Aghaziyarati, Saed Moradi, and Hasan Talebi.
\newblock Small infrared target detection using absolute average difference
  weighted by cumulative directional derivatives.
\newblock {\em Infrared Physics \& Technology}, 101:78--87, 2019.

\bibitem{moradi2020fast}
Saed Moradi, Payman Moallem, and Mohamad~Farzan Sabahi.
\newblock Fast and robust small infrared target detection using absolute
  directional mean difference algorithm.
\newblock {\em Signal Processing}, 177:107727, 2020.

\bibitem{8922738}
Jinhui Han, Saed Moradi, Iman Faramarzi, Chengyin Liu, Honghui Zhang, and Qian
  Zhao.
\newblock A local contrast method for infrared small-target detection utilizing
  a tri-layer window.
\newblock {\em IEEE Geoscience and Remote Sensing Letters}, 17(10):1822--1826,
  2020.

\bibitem{xia2019infrared}
Chaoqun Xia, Xiaorun Li, Liaoying Zhao, and Rui Shu.
\newblock Infrared small target detection based on multiscale local contrast
  measure using local energy factor.
\newblock {\em IEEE Geoscience and Remote Sensing Letters}, 17(1):157--161,
  2019.

\bibitem{zhang2018infrared}
Hong Zhang, Lei Zhang, Ding Yuan, and Hao Chen.
\newblock Infrared small target detection based on local intensity and gradient
  properties.
\newblock {\em Infrared Physics \& Technology}, 89:88--96, 2018.

\bibitem{gao2013infrared}
Chenqiang Gao, Deyu Meng, Yi~Yang, Yongtao Wang, Xiaofang Zhou, and Alexander~G
  Hauptmann.
\newblock Infrared patch-image model for small target detection in a single
  image.
\newblock {\em IEEE Transactions on Image Processing}, 22(12):4996--5009, 2013.

\bibitem{Gao2018Infrared}
Chenqiang Gao, Lan Wang, Yongxing Xiao, Qian Zhao, and Deyu Meng.
\newblock Infrared small-dim target detection based on markov random field
  guided noise modeling.
\newblock {\em Pattern Recognition}, 76(Supplement C):463--475, 2018/04/01/
  2018.

\bibitem{wang2018infrared}
Huan Wang, Fei Yang, Congcong Zhang, and Mingwu Ren.
\newblock Infrared small target detection based on patch image model with local
  and global analysis.
\newblock {\em International Journal of Image and Graphics}, 18(01):1850002,
  2018.

\bibitem{zhang2019infrared1}
Tianfang Zhang, Hao Wu, Yuhan Liu, Lingbing Peng, Chunping Yang, and Zhenming
  Peng.
\newblock Infrared small target detection based on non-convex optimization with
  lp-norm constraint.
\newblock {\em Remote Sensing}, 11(5):559, 2019.

\bibitem{zhang2019infrared2}
Landan Zhang and Zhenming Peng.
\newblock Infrared small target detection based on partial sum of the tensor
  nuclear norm.
\newblock {\em Remote Sensing}, 11(4):382, 2019.

\bibitem{zhao2020novel}
Bin Zhao, Chunping Wang, Qiang Fu, and Zishuo Han.
\newblock A novel pattern for infrared small target detection with generative
  adversarial network.
\newblock {\em IEEE Transactions on Geoscience and Remote Sensing}, 2020.

\bibitem{wang2019miss}
Huan Wang, Luping Zhou, and Lei Wang.
\newblock Miss detection vs. false alarm: Adversarial learning for small object
  segmentation in infrared images.
\newblock In {\em 2019 IEEE/CVF International Conference on Computer Vision
  (ICCV)}, pages 8508--8517. IEEE, 2019.

\bibitem{zhao2019tbc}
Mingxin Zhao, Li~Cheng, Xu~Yang, Peng Feng, Liyuan Liu, and Nanjian Wu.
\newblock Tbc-net: A real-time detector for infrared small target detection
  using semantic constraint.
\newblock {\em arXiv preprint arXiv:2001.05852}, 2019.

\bibitem{shang2016infrared}
Ke~Shang, Xiao Sun, Jinwen Tian, Yansheng Li, and Jiayi Ma.
\newblock Infrared small target detection via line-based reconstruction and
  entropy-induced suppression.
\newblock {\em Infrared Physics \& Technology}, 76:75--81, 2016.

\bibitem{deng2016infrared}
He~Deng, Xianping Sun, Maili Liu, Chaohui Ye, and Xin Zhou.
\newblock Infrared small-target detection using multiscale gray difference
  weighted image entropy.
\newblock {\em IEEE Transactions on Aerospace and Electronic Systems},
  52(1):60--72, 2016.

\bibitem{liu2018tiny}
Jie Liu, Ziqing He, Zuolong Chen, and Lei Shao.
\newblock Tiny and dim infrared target detection based on weighted local
  contrast.
\newblock {\em IEEE Geoscience and Remote Sensing Letters}, 15(11):1780--1784,
  2018.

\bibitem{wu2020double}
Lang Wu, Yong Ma, Fan Fan, Minghui Wu, and Jun Huang.
\newblock A double-neighborhood gradient method for infrared small target
  detection.
\newblock {\em IEEE Geoscience and Remote Sensing Letters}, 2020.

\bibitem{zhou2020robust}
Fei Zhou, Yiquan Wu, Yimian Dai, and Kang Ni.
\newblock Robust infrared small target detection via jointly sparse constraint
  of l1/2-metric and dual-graph regularization.
\newblock {\em Remote Sensing}, 12(12):1963, 2020.

\bibitem{li2018robust}
Yansheng Li and Yongjun Zhang.
\newblock Robust infrared small target detection using local steering kernel
  reconstruction.
\newblock {\em Pattern Recognition}, 77:113--125, 2018.

\bibitem{deng2018adaptive}
Lizhen Deng, Hu~Zhu, Quan Zhou, and Yansheng Li.
\newblock Adaptive top-hat filter based on quantum genetic algorithm for
  infrared small target detection.
\newblock {\em Multimedia Tools and Applications}, 77(9):10539--10551, 2018.

\bibitem{han2020infrared}
Jinhui Han, Chengyin Liu, Yuchun Liu, Zhen Luo, Xiaojian Zhang, and Qifeng Niu.
\newblock Infrared small target detection utilizing the enhanced closest-mean
  background estimation.
\newblock {\em IEEE Journal of Selected Topics in Applied Earth Observations
  and Remote Sensing}, 2020.

\bibitem{deshpande1999max}
Suyog~D Deshpande, Meng~Hwa Er, Ronda Venkateswarlu, and Philip Chan.
\newblock Max-mean and max-median filters for detection of small targets.
\newblock In {\em Signal and Data Processing of Small Targets 1999}, volume
  3809, pages 74--83. International Society for Optics and Photonics, 1999.

\bibitem{gao2008generalised}
Ch-Q Gao, J-W Tian, and Peng Wang.
\newblock Generalised-structure-tensor-based infrared small target detection.
\newblock {\em Electronics Letters}, 44(23):1349--1351, 2008.

\bibitem{chen2013local}
CL~Philip Chen, Hong Li, Yantao Wei, Tian Xia, and Yuan~Yan Tang.
\newblock A local contrast method for small infrared target detection.
\newblock {\em IEEE Transactions on Geoscience and Remote Sensing},
  52(1):574--581, 2013.

\bibitem{deng2016small}
He~Deng, Xianping Sun, Maili Liu, Chaohui Ye, and Xin Zhou.
\newblock Small infrared target detection based on weighted local difference
  measure.
\newblock {\em IEEE Transactions on Geoscience and Remote Sensing},
  54(7):4204--4214, 2016.

\bibitem{shi2017high}
Yafei Shi, Yantao Wei, Huang Yao, Donghui Pan, and Guangrun Xiao.
\newblock High-boost-based multiscale local contrast measure for infrared small
  target detection.
\newblock {\em IEEE Geoscience and Remote Sensing Letters}, 15(1):33--37, 2017.

\bibitem{qin2016effective}
Yao Qin and Biao Li.
\newblock Effective infrared small target detection utilizing a novel local
  contrast method.
\newblock {\em IEEE Geoscience and Remote Sensing Letters}, 13(12):1890--1894,
  2016.

\bibitem{8660588}
Jinhui Han, Sibang Liu, Gang Qin, Qian Zhao, Honghui Zhang, and Nana Li.
\newblock A local contrast method combined with adaptive background estimation
  for infrared small target detection.
\newblock {\em IEEE Geoscience and Remote Sensing Letters}, 16(9):1442--1446,
  2019.

\bibitem{8754801}
Peng Du and Askar Hamdulla.
\newblock Infrared small target detection using homogeneity-weighted local
  contrast measure.
\newblock {\em IEEE Geoscience and Remote Sensing Letters}, 17(3):514--518,
  2020.

\bibitem{8705367}
Yao Qin, Lorenzo Bruzzone, Chengqiang Gao, and Biao Li.
\newblock Infrared small target detection based on facet kernel and random
  walker.
\newblock {\em IEEE Transactions on Geoscience and Remote Sensing},
  57(9):7104--7118, 2019.

\bibitem{han2014robust}
Jinhui Han, Yong Ma, Bo~Zhou, Fan Fan, Kun Liang, and Yu~Fang.
\newblock A robust infrared small target detection algorithm based on human
  visual system.
\newblock {\em IEEE Geoscience and Remote Sensing Letters}, 11(12):2168--2172,
  2014.

\bibitem{wei2016multiscale}
Yantao Wei, Xinge You, and Hong Li.
\newblock Multiscale patch-based contrast measure for small infrared target
  detection.
\newblock {\em Pattern Recognition}, 58:216--226, 2016.

\bibitem{8663277}
Jinyan Gao, Zaiping Lin, and Wei An.
\newblock Infrared small target detection using a temporal variance and spatial
  patch contrast filter.
\newblock {\em IEEE Access}, 7:32217--32226, 2019.

\bibitem{deng2018low}
Xiaoya Deng, Wei Li, Liwei Li, Wenjuan Zhang, and Xia Li.
\newblock Low-rank and sparse decomposition on contrast map for small infrared
  target detection.
\newblock In {\em 2018 24th International Conference on Pattern Recognition
  (ICPR)}, pages 2682--2687. IEEE, 2018.

\bibitem{guan2020infrared}
Xuewei Guan, Landan Zhang, Suqi Huang, and Zhenming Peng.
\newblock Infrared small target detection via non-convex tensor rank surrogate
  joint local contrast energy.
\newblock {\em Remote Sensing}, 12(9):1520, 2020.

\bibitem{8832236}
Wei Li, Mingjing Zhao, Xiaoya Deng, Lu~Li, Liwei Li, and Wenjuan Zhang.
\newblock Infrared small target detection using local and nonlocal spatial
  information.
\newblock {\em IEEE Journal of Selected Topics in Applied Earth Observations
  and Remote Sensing}, 12(9):3677--3689, 2019.

\bibitem{ZHANG201955}
Xiangyue Zhang, Qinghai Ding, Haibo Luo, Bin Hui, Zheng Chang, and Junchao
  Zhang.
\newblock Infrared small target detection based on an image-patch tensor model.
\newblock {\em Infrared Physics \& Technology}, 99:55--63, 2019.

\bibitem{dai2016infrared}
Yimian Dai, Yiquan Wu, and Yu~Song.
\newblock Infrared small target and background separation via column-wise
  weighted robust principal component analysis.
\newblock {\em Infrared Physics \& Technology}, 77:421--430, 2016.

\bibitem{dai2017reweighted}
Yimian Dai and Yiquan Wu.
\newblock Reweighted infrared patch-tensor model with both nonlocal and local
  priors for single-frame small target detection.
\newblock {\em IEEE journal of selected topics in applied earth observations
  and remote sensing}, 10(8):3752--3767, 2017.

\bibitem{wang2017infrared}
Xiaoyang Wang, Zhenming Peng, Dehui Kong, and Yanmin He.
\newblock Infrared dim and small target detection based on stable multisubspace
  learning in heterogeneous scene.
\newblock {\em IEEE Transactions on Geoscience and Remote Sensing},
  55(10):5481--5493, 2017.

\bibitem{simonyan2014very}
Karen Simonyan and Andrew Zisserman.
\newblock Very deep convolutional networks for large-scale image recognition.
\newblock {\em arXiv preprint arXiv:1409.1556}, 2014.

\bibitem{he2016deep}
Kaiming He, Xiangyu Zhang, Shaoqing Ren, and Jian Sun.
\newblock Deep residual learning for image recognition.
\newblock In {\em Proceedings of the IEEE conference on computer vision and
  pattern recognition}, pages 770--778, 2016.

\bibitem{lee2019energy}
Youngwan Lee, Joong-won Hwang, Sangrok Lee, Yuseok Bae, and Jongyoul Park.
\newblock An energy and gpu-computation efficient backbone network for
  real-time object detection.
\newblock In {\em Proceedings of the IEEE/CVF Conference on Computer Vision and
  Pattern Recognition Workshops}, 2019.

\bibitem{xie2016multispectral}
Qi~Xie, Qian Zhao, Deyu Meng, Zongben Xu, Shuhang Gu, Wangmeng Zuo, and Lei
  Zhang.
\newblock Multispectral images denoising by intrinsic tensor sparsity
  regularization.
\newblock In {\em Proceedings of the IEEE conference on computer vision and
  pattern recognition}, pages 1692--1700, 2016.

\bibitem{fan2018dim}
Zunlin Fan, Duyan Bi, Lei Xiong, Shiping Ma, Linyuan He, and Wenshan Ding.
\newblock Dim infrared image enhancement based on convolutional neural network.
\newblock {\em Neurocomputing}, 272:396--404, 2018.

\bibitem{shi2019infrared}
Manshu Shi and Huan Wang.
\newblock Infrared dim and small target detection based on denoising
  autoencoder network.
\newblock {\em Mobile Networks and Applications}, pages 1--15, 2019.

\bibitem{ju2021istdet}
Moran Ju, Jiangning Luo, Guangqi Liu, and Haibo Luo.
\newblock Istdet: An efficient end-to-end neural network for infrared small
  target detection.
\newblock {\em Infrared Physics \& Technology}, 114:103659, 2021.

\bibitem{hou2021ristdnet}
Qingyu Hou, Zhipeng Wang, Fanjiao Tan, Ye~Zhao, Haoliang Zheng, and Wei Zhang.
\newblock Ristdnet: Robust infrared small target detection network.
\newblock {\em IEEE Geoscience and Remote Sensing Letters}, 2021.

\bibitem{ryu2019heterogeneous}
Junhwan Ryu and Sungho Kim.
\newblock Heterogeneous gray-temperature fusion-based deep learning
  architecture for far infrared small target detection.
\newblock {\em Journal of Sensors}, 2019, 2019.

\bibitem{gao2019dim}
Zhisheng Gao, Jiao Dai, and Chunzhi Xie.
\newblock Dim and small target detection based on feature mapping neural
  networks.
\newblock {\em Journal of Visual Communication and Image Representation},
  62:206--216, 2019.

\bibitem{du2021cnn}
Jinming Du, Huanzhang Lu, Moufa Hu, Luping Zhang, and Xinglin Shen.
\newblock Cnn-based infrared dim small target detection algorithm using
  target-oriented shallow-deep features and effective small anchor.
\newblock {\em IET Image Processing}, 15(1):1--15, 2021.

\bibitem{7298594}
Christian Szegedy, Wei Liu, Yangqing Jia, Pierre Sermanet, Scott Reed, Dragomir
  Anguelov, Dumitru Erhan, Vincent Vanhoucke, and Andrew Rabinovich.
\newblock Going deeper with convolutions.
\newblock In {\em 2015 IEEE Conference on Computer Vision and Pattern
  Recognition (CVPR)}, pages 1--9, 2015.

\bibitem{zhu2020tnlrs}
Hu~Zhu, Haopeng Ni, Shiming Liu, Guoxia Xu, and Lizhen Deng.
\newblock Tnlrs: Target-aware non-local low-rank modeling with saliency
  filtering regularization for infrared small target detection.
\newblock {\em IEEE Transactions on Image Processing}, 29:9546--9558, 2020.

\bibitem{dai21acm}
Yimian Dai, Yiquan Wu, Fei Zhou, and Kobus Barnard.
\newblock Asymmetric contextual modulation for infrared small target detection.
\newblock {IEEE} Winter Conference on Applications of Computer Vision, {WACV},
  2021.

\bibitem{kingma2014adam}
Diederik~P Kingma and Jimmy Ba.
\newblock Adam: A method for stochastic optimization.
\newblock {\em arXiv preprint arXiv:1412.6980}, 2014.

\end{thebibliography}

\end{document}